\documentclass[journal, 12pt, draftclsnofoot, onecolumn]{IEEEtran}
\IEEEoverridecommandlockouts

\usepackage{amsmath,amssymb,amsfonts,steinmetz}
\usepackage{mathtools,algpseudocode,algorithm,MnSymbol}

\usepackage{bm}
\usepackage[pdftex]{graphicx}
\usepackage{textcomp}
\usepackage{color}
\usepackage{longtable}
\usepackage{gensymb}
\usepackage{booktabs}
\usepackage{indentfirst}   
\usepackage{array}
\usepackage{multirow}
\usepackage{caption, subcaption}
\usepackage{tikz}
\usepackage{tabularx,ragged2e}
\usepackage{float}
\usepackage[inline]{enumitem}
\usepackage{nicefrac}
\usepackage{balance}
\usepackage{url}
\usepackage{pgfplots}
\usetikzlibrary{shapes.multipart,intersections}
\usepackage{cite}
\usepackage{xcolor}
\def\BibTeX{{\rm B\kern-.05em{\sc i\kern-.025em b}\kern-.08em
    T\kern-.1667em\lower.7ex\hbox{E}\kern-.125emX}}
\usetikzlibrary{arrows.meta}
\usetikzlibrary{calc}
\makeatletter
\usepackage[nolist]{acronym}
\makeatother
\usepackage[draft]{todonotes}   
\usepackage{upgreek}
\usepackage{seqsplit}
\usepackage[font={footnotesize},hypcap={false}]{caption}
\usepackage{hyperref}
\usepackage[utf8]{inputenc} 
\acrodef{RIS}{reconfigurable intelligent surface}
\acrodef{BS}{base station}
\acrodef{FF}{far field}
\acrodef{UE}{user equipment}
\acrodef{LoS}{line-of-sight}
\acrodef{NLoS}{non-line-of-sight}
\acrodef{NF}{nearfield}
\acrodef{SNR}{signal-to-noise ratio}
\acrodef{SINR}{signal-to-interference-and-noise-ratio}
\acrodef{SISO}{single-input-single-output}
\acrodef{PEB}{position error bound}
\acrodef{FIM}{Fisher information matrix}
\acrodef{SDP}{semidefinite program}
\acrodef{PSD}{positive semidefinite}
\acrodef{LMI}{linear matrix inequality}
\acrodef{MC}{multi-carrier}
\acrodef{MIMO}{multiple inputs multiple outputs}
\acrodef{OEB}{orientation error bound}
\acrodef{DoD}{Direction of Departure}
\acrodef{TDoA}{Time Difference of Arrival}
\acrodef{RIS}{Reconfigurable intelligent surface}
\acrodef{w.r.t.}{with respect to}
\acrodef{SRE}{Smart radio environment}
\acrodef{TX}{transmitter}
\acrodef{RX}{receiver}
\acrodef{QoS}{Quality of Service}
\acrodef{DFT}{Discrete Fourier Transform}

\newcommand{\h}{\mathsf{H}}

\newcommand{\pp}{\boldsymbol{p}}
\renewcommand{\aa}{\boldsymbol{a}}
\renewcommand{\AA}{\boldsymbol{A}}
\newcommand{\bb}{\boldsymbol{b}}
\newcommand{\BB}{\boldsymbol{B}}
\newcommand{\CC}{\boldsymbol{C}}
\newcommand{\xx}{\boldsymbol{x}}
\newcommand{\XX}{\boldsymbol{X}}
\newcommand{\yy}{\boldsymbol{y}}
\newcommand{\JJ}{\boldsymbol{J}}
\renewcommand{\gg}{\boldsymbol{g}}
\newcommand{\ff}{\boldsymbol{f}}
\newcommand{\FF}{\boldsymbol{F}}
\newcommand{\nn}{\boldsymbol{n}}
\newcommand{\II}{\boldsymbol{I}}
\newcommand{\uu}{\boldsymbol{u}}
\newcommand{\UU}{\boldsymbol{U}}
\newcommand{\ee}{\boldsymbol{e}}
\newcommand{\boldone}{\boldsymbol{1}}

\newcommand{\pprx}{\pp_{\text{\ac{RX}}}}
\newcommand{\pptx}{\pp_{\text{\ac{TX}}}}

\newcommand{\omegab}{\bm{\omega}}
\newcommand{\Omegab}{\bm{\Omega}}
\newcommand{\upzetab}{\bm{\upzeta}}
\newcommand{\upmub}{\bm{\upmu}}
\newcommand{\lambdab}{\bm{\lambda}}
\newcommand{\Lambdab}{\bm{\Lambda}}

\newcommand{\gettikzxy}[3]{%
  \tikz@scan@one@point\pgfutil@firstofone#1\relax
  \edef#2{\the\pgf@x}%
  \edef#3{\the\pgf@y}%
}

\begin{document}


\title{Performance of RIS-Aided Nearfield Localization under Beams Approximation from Real Hardware Characterization}

\author{\IEEEauthorblockN{Moustafa Rahal\IEEEauthorrefmark{1}\IEEEauthorrefmark{3}, Beno\^{i}t Denis\IEEEauthorrefmark{1}, Kamran Keykhosravi\IEEEauthorrefmark{2}, \\Musa Furkan Keskin\IEEEauthorrefmark{2}, Bernard Uguen\IEEEauthorrefmark{3}, George C.~Alexandropoulos\IEEEauthorrefmark{4}, and Henk Wymeersch\IEEEauthorrefmark{2}}\\
\IEEEauthorblockA{\IEEEauthorrefmark{1}CEA-Leti, Université Grenoble Alpes, F-38000 Grenoble, France\\
\IEEEauthorrefmark{2}Department of Electrical Engineering, Chalmers University of Technology, Gothenburg, Sweden\\
\IEEEauthorrefmark{3}Université Rennes 1, IETR - UMR 6164, F-35000 Rennes, France\\
}
\IEEEauthorrefmark{4}Department of Informatics and Telecommunications,
National and Kapodistrian University of Athens, Greece
\thanks{Correspondent author: Moustafa Rahal (contact: moustafa.rahal@cea.fr).}
}
\maketitle

\begin{abstract}
The technology of reconfigurable intelligent surfaces (RIS) has been showing promising potential in a variety of applications relying on Beyond-5G networks. \ac{RIS} can indeed provide fine channel flexibility to improve communication quality of service (QoS) or restore localization capabilities in challenging operating conditions, while conventional approaches fail (e.g., due to insufficient infrastructure, severe radio obstructions). In this paper, we tackle a general low-complexity approach for optimizing the precoders that control such reflective surfaces under hardware constraints. More specifically, it allows the approximation of any desired beam pattern using a pre-characterized look-up table of feasible complex reflection coefficients for each \ac{RIS} element. The proposed method is first evaluated in terms of beam fidelity for several examples of \ac{RIS} hardware prototypes. Then, by means of a theoretical bounds analysis, we examine the impact of \ac{RIS} beams approximation on the performance of near-field downlink positioning in non-line-of-sight conditions, while considering several \ac{RIS} phase profiles (incl. directional, random and localization-optimal designs). 
Simulation results in a canonical scenario illustrate how the introduced \ac{RIS} profile optimization scheme can reliably produce the desired \ac{RIS} beams under realistic hardware limitations. They also highlight its sensitivity to both the underlying hardware characteristics and the required beam kinds in relation to the specificity of RIS-aided localization applications.
\end{abstract}



\section*{Introduction}

\acp{RIS} consist of (semi-)passive controllable electromagnetic mirrors or sensing surfaces, and have been identified as a key enabling technology for the upcoming sixth (6G) generation of wireless systems \cite{huang2019reconfigurable,qignqingwu2019}. They are indeed expected to reinforce both the service continuity and the \ac{QoS} of communication networks, or even to locally boost their performance on demand, while limiting the need for additional costly elements of infrastructure (e.g., active \ac{BS}). Although they were mostly intended to extend coverage under severe radio blockages, they have also shown fine capabilities to purposely shape the wireless propagation channel in a variety of location-dependent applications, such as \ac{UE} localization (in both geometric \ac{NF} and \ac{FF} regimes), physical environment mapping and distributed spectrum sensing, or limitation of unintentional radio emissions (e.g., for improved communication security or reduced field exposure) \cite{RISE6G_COMMAG,Kamran_Leveraging}.

 
The different unit elements of a \ac{RIS} can be optimized, for instance in order to concentrate the reflected energy in the \ac{UE} direction, similar to phased array systems \cite{Molisch_HBF_2017_all}. Many models have already been proposed accordingly, including phase control \cite{huang2019reconfigurable}, quantized phase control \cite{alexandg_2021}, amplitude-dependent phase control \cite{Abeywickrama_2020}, and joint amplitude and phase control \cite{Larsson_2021}, which all require a dedicated optimization procedure.
However, as \acp{RIS} are usually made of low-cost hardware, they may be naturally subject to imperfections, impairments, non-linearity effects, dispersed characteristics... \cite{RIS_Impairments,shen2020beamforming}. If not properly taken into account, these phenomena can significantly alter the result of the \ac{RIS} optimization and hence ultimately, the \ac{RIS} behavior itself, compared to the idealized case (i.e., with an unconstrained \ac{RIS} profile).
One unified way to address the various control models above while accounting for hardware characteristics is to use a lookup table. For a reflective RIS, the latter typically contains all the feasible complex reflection coefficients at each of the surface elements (i.e., in terms of amplitude and phase) for a certain \ac{RIS} hardware. In addition, such lookup tables can account for extra coupling effects, which are very complex to model analytically otherwise.

In this paper, we present a computationally efficient method initially introduced in \cite{Rahal_EuCNC22}, which can optimize the configuration of a reflective \ac{RIS} from an a priori imposed lookup table 
with the aim of approximating complex beam patterns.
%
Although the previous approach is generic and applicable to any type of \acp{RIS} or beams, 
as a concrete and practical study case, we herein consider applying it into the specific context of RIS-aided \ac{NF} localization, which somehow requires combining particular \ac{RIS} beams to reach optimality.
Localization-optimal \ac{RIS} phase configurations can indeed be determined through \ac{PEB} optimization, assuming prior information about the \ac{UE} \cite{Elzanaty2021,Rahal_Localization-Optimal_RIS}. 
The aim is hence to improve positioning performance in \ac{NLoS} situations, while relying on a unique \ac{RIS}-reflected path (i.e., as estimated at the \ac{UE}) over a sequence of \ac{SISO} downlink transmissions \cite{keykhosravi2021siso,rahal2021ris} (See Fig.~\ref{fig:Geometry}). Aiming more specifically at improving \ac{NF} localization performance in \cite{Rahal_Localization-Optimal_RIS}, it has been shown that such optimization would necessitate the use of four distinct types of beams at the reflective \ac{RIS} (i.e., one steering beam and its three derivatives \ac{w.r.t.} spherical coordinates). Those four desired beams will be taken here as references into the proposed synthesis process, and compared to directional or random designs. 

Overall, the main paper contributions are as follows: \emph{(i)} Leveraging the recent results from \cite{Rahal_EuCNC22} and \cite{Rahal_Localization-Optimal_RIS} recalled above, we apply a low-complexity \ac{RIS} beams approximation technique to RIS-aided \ac{NLoS} localization with several a priori \ac{RIS} design assumptions (incl. random, directional and localization-optimal options); \emph{(ii)} Accordingly, our analysis now includes not only a qualitative assessment of the fidelity of approximated \ac{RIS} beams \ac{w.r.t.} the desired beams (by means of a 1D/2D beam shape analysis), but also a quantitative assessment of the practical performance degradation induced by such beams approximation in comparison with their respective unconstrained  configurations (by means of a \ac{PEB} analysis); \emph{(iii)} Regarding input hardware constraints, we also extend the use of look-up tables (resulting from experimental characterization) to that of four distinct real \ac{RIS} prototypes \cite{fara2021prototype,DiPalma_2017,GNW_Prototype} and we devise their impact on localization performance in light of their respective design specificity.
\subsubsection*{Notations}
Vectors and matrices are, respectively, denoted by lower-case and upper-case bold letters (e.g., $\xx, \XX$). The notation $[\aa]_{i}$ is used to point at the $i$-th element of vector $\aa$, and similarly, $[\AA]_{i,j}$ represents the element in the $i$-th row and $j$-th column of matrix $\AA$, while $i:j$ is used to specify all the elements between indices $i$ and $j$. The Hadamard product is denoted by $\odot$ and $\dot{\aa}_{x}=\partial \aa/ \partial x$ is the partial derivation of $\aa$ \ac{w.r.t.} $x$. Moreover, the notations $(.)^\top$, $(.)^{*}$ and $(.)^{\mathsf{H}}$ denote the matrix transposition, conjugation, and Hermitian conjugation, respectively. Finally, $(.)^{(r)}$ denotes the $r$-th iteration in a loop and $\text{proj}_{\mathcal{V}}(\xx)$ represents the projection of vector $\xx$ onto a set $\mathcal{V}$.
The operator $\mathrm{tr}(\XX)$ denotes the trace of matrix $\XX$ and $\text{diag}(\xx)$ denotes a diagonal matrix with diagonal elements defined by vector $\xx$.  Finally, $\Vert\cdot\Vert$ is the $l_{2}$-norm operator and $(.)^\smallstar$ denotes the solution of an optimization problem.

\section*{Related State of the Art}

The optimization of \ac{RIS} profiles
(i.e., the fine tuning of their element-wise reflection coefficients and/or phases) is a particularly challenging task in itself which can be seen as a beam design problem. This has been extensively studied for phased arrays, in both communication and radar literature \cite{monopulse_review}. 
%
%
The main difference, however, between a reflective \ac{RIS} and a phased array radar is that in order to create azimuth- and elevation-difference beams in reception via the latter, separate beamforming architectures are commonly employed  \cite{phasedArray_99,phasedArray_2016}; while on the other hand, a reflective \ac{RIS} is only able to passively reflect the impinging signals through optimized phase control and does not produce receive beams via dedicated hardware. 

%
%
In a pure \ac{RIS}-aided communication context, the usual approach is to aim at increasing \ac{SNR} at the receiver, relying on the estimated cascaded channel responses (i.e., between \ac{BS} and \ac{RIS}, and between \ac{RIS} and \ac{UE}), or possibly simply based on the prior \ac{UE} location information (exact or more likely just estimated)~\cite{abrardo2020intelligent}.
The corresponding problem 
can hence be solved out by maximizing recovery performance, 
based on the \ac{DFT} 
and Hadamard matrices,
or by leveraging location information from an external system~\cite{hu2020location}.
%

As for RIS-aided localization specifically, a variety of control solutions have also been put forward recently, depending on the corresponding positioning strategy and operating context. In \cite{wymeersch_beyond_2020} for instance, based on prior \ac{UE} location, simple directional reflection beams are applied at a few \acp{RIS}, which are down-selected to maximize the amount of available location information (in a Fisher Information sense) while avoiding multipath interference. 
In \cite{keykhosravi2021siso}, even simpler random \ac{RIS} phase profiles are used to jointly enable downlink \ac{SISO} positioning and synchronization, relying on \ac{MC} transmissions, without necessitating prior information.
In \cite{Elzanaty2021}, RIS
phases and beamformers are jointly optimized with respect to both the \ac{PEB} and the \ac{OEB} in a generic \ac{MIMO} \ac{MC} context. To overcome both the high number of real optimization variables and the problem non-convexity (resulting from the joint optimization of coupled variables), a simpler approach is then proposed that maximizes the sum of the \acp{SNR} for the central subcarrier at each \ac{BS} antenna. An upper bound on the \ac{SNR} is first maximized, before applying additional constant phase shifts to the \ac{RIS} profile so that both direct and reflective paths are quasi-coherently summed up at each receiving antenna.

Focusing on the specific case of \ac{NF} localization, many other \ac{RIS} optimization schemes have been put forward, such as \ac{SNR} maximization in \cite{Rinchi22} and \cite{Palmucci22} and \ac{DFT} matrices in \cite{Pan22}. On the other hand, a traditional randomized \ac{RIS} phase design was considered in \cite{Ghazalian22} to achieve bi-static localization of the \ac{RIS}. In \cite{rahal2021ris}, random profiles are also utilized and the authors showed, theoretically, that this profile design can ensure localization continuity in \ac{NLoS} conditions through direct positioning out of one single RIS-reflected path, relying on \ac{NF} properties. However random profiles are still clearly suboptimal in terms of localization performance. In \cite{Rahal_Localization-Optimal_RIS}, based on \ac{PEB} optimization in the same \ac{NF} \ac{NLoS} context, it is shown that four distinct beams must be applied (i.e., one steering beam and its three derivatives as a function of spherical coordinates), while adjusting their relative weights (in the power or time domain, indifferently) depending on prior \ac{UE} location. Finally, in \cite{Mei22}, \ac{RIS} phase design and quantization are ruled by hardware and its effect is studied on the \ac{NF} focusing performance, but no practical constrained beam generation method is proposed.

Additionally, various works have considered beam design and optimization as \cite{alexandropoulos_near-field_2022} which presents a fast alignment algorithm for the phase shifts of the \ac{RIS} and the transceiver beam-former, which uses a variable-width hierarchical phase-shift codebook. Another recent study \cite{liu_low-overhead_2022} proposes a far-field beam-based beam training scheme, where a deep residual network is utilized to estimate the optimal near-field \ac{RIS} codeword. 
In \cite{chung_location-aware_2021}, the authors present a beamwidth adaptation technique during the beam training process, and an atomic norm channel estimation method for the cascaded channel.
 The paper \cite{chen-hu_differential_2022} proposes the use of differential data to determine the best beam for the \ac{RIS}. 
Furthermore, \cite{palmucci_ris-aided_2022} explores a joint design of the reflection coefficients of multiple \acp{RIS} and the precoding strategy of a single \ac{BS} to optimize the tracking of the position and velocity of a single \ac{UE} with an iterative block coordinate descent algorithm.

One major difficulty while optimizing \ac{RIS} beams under such hardware constraints is that the latter may be difficult to handle analytically. As an example, unit-norm constraints are usually non-convex and shall necessitate specific iterative approaches \cite{tranter2017}. In some other cases, the addressable \ac{RIS} configurations may be quantized \cite{DiPalma_2017}, subject to phase-dependent amplitude variations (wanted or unwanted) \cite{Abeywickrama_2020}, or only characterized experimentally as a lookup table derived from real measurements \cite{fara2021prototype}.
Under realistic hardware limitations, \ac{RIS} profile optimization can hence be performed either \textit{i}) by directly optimizing the constrained \ac{RIS} configuration with respect to the considered objective (e.g., minimize the \ac{PEB} in the localization context) \cite{huang2019reconfigurable,qignqingwu2019,Abeywickrama_2020, Huang_GLOBECOM_2019} or \textit{ii}) by carrying out an unconstrained optimization of the \ac{RIS} configuration first and then determining the best approximation that could be practically supported by the \ac{RIS}. These two options can be referred to as \emph{``constrain, then optimize''} or \emph{``optimize, then constrain''} (e.g., \cite{Rahal_Localization-Optimal_RIS}), respectively. For the former, low-complexity numerical methods offering good fidelity with respect to the desired (i.e., unconstrained) beam patterns are still needed, which is one of the purposes of this paper.
%
 

\section*{System Model and General Problems Formulation}
%

\subsection*{Overall System Model and Localization Scenario} 
Following \cite{basar2019wireless}, we first consider a single-antenna \ac{TX}, which is communicating in \ac{NLoS} with a single-antenna \ac{RX}. The transmission is hence enabled through an $M$-element reflective RIS, over two \ac{LoS} links (i.e., \ac{TX}-\ac{RIS} and \ac{RIS}-\ac{RX} links). In case of narrowband transmissions, and static and known RIS location, the corresponding baseband received signal is then expressed as follows \cite{rahal2021ris,AbuShaban_2021}:
\begin{equation}
\begin{split}
y & =\alpha \aa^{\top}(\pprx)\Omegab\aa(\pptx) x + n \\
&= \alpha \omegab^\top \bb(\pprx,\pptx) x  + n,\label{eq:Received_signal}
\end{split}
\end{equation}
where $\pptx$ and $\pprx$ are respectively the \ac{TX} and \ac{RX} positions, $\alpha$ is the complex channel gain, $\Omegab\triangleq{\rm diag}(\omegab)$ with $\omegab\in \mathbb{C}^{M\times 1}$ accounting for the \ac{RIS} complex configuration, $x$ is the transmitted signal of energy $E_{\text{s}}$,
$n\sim\mathcal{CN}(0,N_0)$ is the additive white Gaussian noise of power spectral density $N_0$, and $\aa(\cdot)\in \mathbb{C}^{M\times 1}$ indicates the \ac{RIS} response. Considering a point $\pp$, the $m$-th ($m=1,2,\ldots,M$) entry of $\aa(\pp)$, corresponding to the $m$-th \ac{RIS} element located at $\pp_{m}$, is given by\footnote{Note that the model converges to the standard far-field model when the distance $\Vert \pp-\pp_{\text{RIS}}\Vert$ becomes large.} 
\begin{align}
    [\aa(\pp)]_{m}=\exp\left(-\jmath\frac{2\pi}{\lambda}\left(\Vert\pp-\pp_{m}\Vert - \Vert\pp-\pp_{\text{RIS}}\Vert \right)\right),\label{eq:RIS_response}
\end{align}
where $\pp_{\text{RIS}}$ is the \ac{RIS} center location and $\lambda$ is the wavelength. Finally, in \eqref{eq:Received_signal}, we have used the definition $\bb(\pprx,\pptx)\triangleq{\aa}(\pprx) \odot \aa(\pptx)$.
\begin{figure}[ht]
 \centering
 \includegraphics[width=0.8\linewidth]{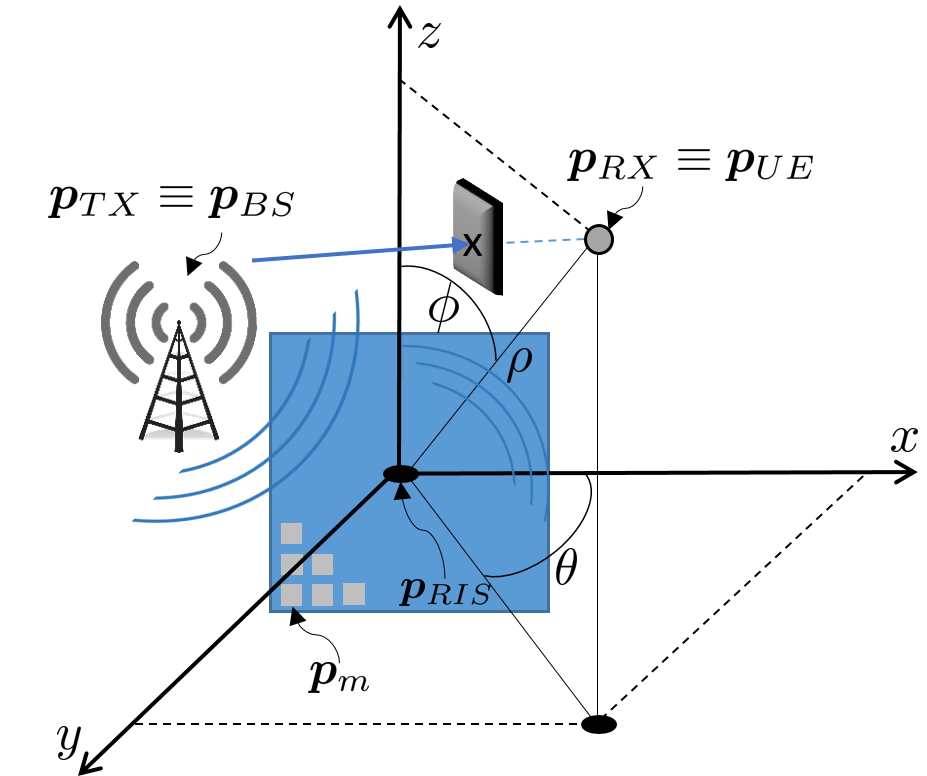}
 \caption{Typical \ac{NF} \ac{NLoS} positioning scenario over \acl{SISO} downlink transmissions with one single reflective \ac{RIS} and related problem geometry with respect to the \ac{RX} point $\pp$, where the \ac{RIS} center serves as the origin of both the spherical and Cartesian coordinates systems.}
 \label{fig:Geometry}
\end{figure}
Fig.~\ref{fig:Geometry} illustrates a millimeter wave (mmWave) downlink \ac{NLoS} localization scenario, where a \ac{BS} in $\pptx=\pp_{\text{\ac{BS}}}$ broadcasts a narrowband pilot signal $x_{t} \in \mathbb{C}$ over $T$ transmissions, with a bandwidth $W$, a transmit power $P_{\text{tx}}$, and a total transmit energy $E_{\text{tot}} = E_{s}MT=\nicefrac{(P_{\text{tx}}MT)}{W}$. From \eqref{eq:Received_signal}, the complex signal $y_t \in \mathbb{C}$ received by a \ac{UE} in $\pprx=\pp_{\text{UE}}$ at time $t$ after \ac{RIS} reflection can be written as
\vspace{-1mm}
\begin{align}
    y_{t} & =\alpha \aa^{\top}(\pp_{\text{UE}})\Omegab_{t}\aa(\pp_{\text{BS}}) x_{t}  
    + n_{t}, \label{eq:Received_signal_time}
\end{align}
where $\Omegab_{t} = \text{diag}(\omegab_{t}) $ with $\omegab_{t} \in \mathbb{C}^{M\times 1}$, which can vary as a function of time. 
This received signal can also be vectorized over the $T$ transmissions into $\yy \in \mathbb{C}^{T\times1}$ as follows\footnote{Without loss of generality, we assume a constant pilot $x_t=\sqrt{E_s}$ is transmitted.}
\vspace{-3mm}
\begin{align}
    \yy=\sqrt{E_{s}}\alpha \FF^{\top} \aa(\pp_{\text{UE}}) + \nn, \label{eq:generalized_received_signal}
\end{align}
where $\FF= [\ff_1,\ldots,\ff_T] \in \mathbb{C}^{M\times T}$ with $\ff_t = \Omegab_t \aa(\pp_{\text{BS}}) \in \mathbb{C}^{M\times1}$.

In the absence of \ac{LoS} and at reasonably short \ac{RIS}-\ac{UE} distances and/or with large surfaces, one can leverage properties of the \ac{NF} \ac{RIS} response (i.e., location-dependent information conveyed by radio wavefront curvature) to estimate the position of the \ac{UE} based on the observed received signals over time. Here, it is assumed that the position of the \ac{TX} and the \ac{RIS}, the orientation of the \ac{RIS},  and the \ac{RIS} profiles are known, where the \ac{RIS} profiles may be optimized to improve the accuracy of estimation.


\subsection*{\ac{RIS} Beam Approximation Problem}
Due to \ac{RIS} hardware limitations and/or design specificity~\cite{alexandg_2021}, all element-wise reflection coefficients are assumed to take values among a few discrete complex values only. In other words, for the $m$-th element of the vector $\omegab$ in \eqref{eq:Received_signal}, it holds that $\omega_m \in \mathcal{V}$, where $\mathcal{V}$ is a finite set of complex numbers, whose magnitude cannot exceed unity as the \ac{RIS} is passive. For instance in \cite[Table I]{fara2021prototype}, for a particular hardware prototype, the individual \ac{RIS} element response has been experimentally characterized, and the corresponding set $\mathcal{V}$ has the cardinality of 14. Leveraging the model in \eqref{eq:Received_signal}, we  first tackle a generic optimization framework that aims at approximating any desired beam pattern for a reflective RIS, i.e.
\begin{align}\label{eq:beam_pattern}
    G(\pp)= \omegab^\top\bb(\pp,\pptx)
\end{align}
for an arbitrary \ac{RX} point $\pp$ in the coverage area $\mathcal{G}$. Without loss of generality, 
$\pptx$ is assumed to be static.

As already alluded in the previous sections, concrete examples of usual canonical beam patterns to be synthesized in RIS-aided applications 
include \emph{steering} beams (such as DFT beams)~\cite{AbuShaban_2021}, \emph{derivative} beams (e.g., difference beams just like in monopulse radar~\cite{monopulse_review}, MIMO radar~\cite{li2007range}, or even localization~\cite{keskin_optimal_2021}), or even \emph{multiple concurrent} beams~\cite{zhang2018multibeam,Rahal_EuCNC22}.
%
%
As seen in the following, localization-optimal RIS configurations may request a combination of the latter canonical beams, jointly or sequentially depending on the implementation (typically, a steering beam pointing to the \ac{UE}, and its derivatives \ac{w.r.t.} the three spherical coordinates).
So one more goal here is to assess the theoretical performance degradation induced by such beam approximations directly at the application level (i.e., typically, in terms of \ac{PEB}), beyond beam fidelity issues.

\section*{Proposed RIS Beam Synthesis Methodology}
%
%
\subsection*{Least Squares Precoder Design}
Leveraging the approach in \cite{tranter2017}, the desired beam pattern in \eqref{eq:beam_pattern} is first discretized into a $N_G$-element vector $\gg$. Each element of $\gg$ corresponds to a specific point in $\mathcal{G}$. $N_G$ hence refers to the number of such discrete locations sampled from $\mathcal{G}$. Accordingly, it comes that $[\gg]_k=G(\pp_k)$, with $k=1,2,\ldots,N_G$. After defining the $N_G \times M$ complex-valued matrix $\BB\triangleq[\bb^\top(\pp_1,\pptx);\ldots;\bb^\top(\pp_{N_G}, \pptx)]$, 
the \ac{RIS} configuration optimization problem can be formulated as follows, like in \cite[eq.~(12)]{tranter2017}: 
\begin{subequations}
\label{eq:OPT1}
\begin{align}
    \min_{s,\omegab} & \quad \Vert \gg - s \BB \omegab\Vert^2\\ \label{eq_v_cons}
    \text{s.t.} &\quad  \omega_m \in \mathcal{V},\,m=1,2,\ldots, M,
\end{align} 
\end{subequations}
where $s \in \mathbb{C}$ is a normalization factor enabling to solve the scaling issue for $\gg$ while designing the beam pattern~\cite{tranter2017}. The optimization approach in \eqref{eq:OPT1} represents a RIS beam pattern synthesis problem aiming to determine the optimal RIS phase profile $\omegab$ that best matches a given desired beam pattern $\gg$.

To deal with the \ac{RIS} configuration problem exposed above, 
a simple projected gradient descent algorithm is used (See Algorithm~\ref{alg:algo1}), inheriting from \cite[Alg.~2]{tranter2017}. First of all, one gradient descent step is performed \ac{w.r.t.} scaling variable $s$ (See Line\,\ref{A1:L2} of Algorithm~\ref{alg:algo1}).
Then, we apply an unconstrained gradient descent step \ac{w.r.t.} \ac{RIS} configuration (See Line\,\ref{A1:L3}), whose result is subsequently projected onto the set $\mathcal{V}$ (See Line\,\ref{A1:L4}) so as to satisfy the look-up table constraint in \eqref{eq_v_cons}. Note that $\beta$ is just a design parameter that enables to adjust the step size, while $\lambda_{\max}(\cdot)$ refers to the largest eigenvalue of the matrix argument. Algorithm~\ref{alg:algo2} summarizes the projection step. Accordingly, the complexity of Algorithm~\ref{alg:algo1} (per iteration) is $\mathcal{O}(M (N_G + \lvert \mathcal{V} \rvert))$.
\begin{algorithm}[t]
\caption{RIS Configuration Design}
\label{alg:algo1}
\textbf{Initialize:} $\beta\in (0,1)$ and $\omegab^{(0)}=\text{proj}_{\mathcal{V}}(\BB^\dagger \gg)$.
\begin{algorithmic}[1]
    \For {$r=1,2,\ldots$} 
        \State Compute $s^{(r)}=\frac{(\omegab^{(r-1)})^{\h}\BB^{\h}\gg}{\Vert\BB\omegab^{(r-1)} \Vert^2}$. \label{A1:L2}
        \vspace{2mm}
        \State Set $\omegab^{(r)}_{\text{u}}=\omegab^{(r-1)}+\frac{\beta \BB^{\h}(\gg-s^{(r)}\BB\omegab^{(r-1)})}{\lambda_{\text{max}}(|s^{(r)}|^{2}\BB^{\h}\BB)}$.\label{A1:L3}
        \vspace{2mm}
        \State Calculate $\omegab^{(r)}=\text{proj}_{\mathcal{V}}(\omegab^{(r)}_{\text{u}})$ using Algorithm~\ref{alg:algo2}.\label{A1:L4}
    \EndFor
\end{algorithmic}
\end{algorithm}
\begin{algorithm}[t]
\caption{Projection onto set $\mathcal{V}$: $\omegab_{\text{out}}=\text{proj}_{\mathcal{V}}(\omegab_{\text{in}})$}
\label{alg:algo2}
\begin{algorithmic}[1]
\For {$m=1,\ldots,M$}
\State Find $[{\omega}_{\text{out}}]_m=\arg \min_{{\omega} \in \mathcal{V}} |[\omegab_{\text{in}}]_m-\omega|^2$.
\EndFor
\end{algorithmic}
\end{algorithm}
\subsection*{Reduced-Complexity Solution}
As the definition domain of $G(\pp)$ may coincide with the full 3D space, $N_G$ can become tremendously large, and even prohibitive, resulting in very high computational complexity for Algorithm~\ref{alg:algo1}. To relax this burden, one can express the beams at stake in terms of spherical coordinates and re-define accordingly the optimization objective from \eqref{eq:OPT1}, with $\rho$ the distance, $\theta$ the azimuth angle and $\phi$ the elevation angle, all defined \ac{w.r.t.} the \ac{RIS} coordinates system (see Fig.~\ref{fig:Geometry}), 
like in \cite[Fig.1b]{keykhosravi2021siso}.


%

Let $\pp_{\text{ref}}$ be the reference \ac{RX} position in spherical coordinates, expressed as  $[\rho_{\text{ref}},\theta_{\text{ref}},\phi_{\text{ref}}]^\top$. Then, we define the following three beam patterns each time varying only one of the spherical dimensions (instead of covering the entire 3-D volume): 
\begin{align}
    \gg_{\rho}& =G([\rho,\theta_{\text{ref}},\phi_{\text{ref}}]^\top),~\rho \in \mathcal{R},\\
    \gg_{\theta}& =G([\rho_{\text{ref}},\theta,\phi_{\text{ref}}]^\top),~\theta \in \mathcal{T},\\
    \gg_{\phi}& =G([\rho_{\text{ref}},\theta_{\text{ref}},\phi]^\top),~\phi \in \mathcal{P},
\end{align}
where $\mathcal{R}$, $\mathcal{T}$, and $\mathcal{P}$ corresponds to the discrete sets of respective spherical coordinates.

Likewise, the different \ac{RIS} response vectors are also re-defined, e.g., \ac{w.r.t.} the azimuth: $\BB_{\theta}=[\bb^\top_{\theta,1};\ldots;\bb^\top_{\theta,|\mathcal{T}|}]$ where  $\bb^\top_{\theta,k}=\bb([\rho_{\text{ref}},\theta_k,\phi_{\text{ref}}]^\top)$ for $\theta_k$ being the $k$-th element ($k=1,2,\ldots,|\mathcal{T}|$) in $\mathcal{T}$.
All in all, the optimization problem can now be re-formulated as follows:
\begin{subequations}\label{eq_prob_lowcomp}
\begin{align}
    \min_{s,\omegab} & \quad \sum_{\text{p} \in \{ \rho,\theta,\phi\}}\Vert \gg_{\text{p}} - s \BB_{\text{p}} \omegab\Vert^2\\
    \text{s.t.} &\quad  \omega_m \in \mathcal{V}, m=1,\ldots, M.
\end{align} 
\end{subequations}
Just like for Algorithm\,\ref{alg:algo1}, the optimization problem above is solved through a gradient-descent step \ac{w.r.t.} the scaling factor $s$ (See Line\,2 of Algorithm\,\ref{alg_lowcomp}) and the vector of RIS phase shifts $\omegab$ (See Line\,3 of Algorithm\,\ref{alg_lowcomp}), before projecting finally the updated vector into the space $\mathcal{V}$ (See Line\,4 of Algorithm\,\ref{alg_lowcomp}). In this case, it is worth noting that $|\mathcal{R}|+|\mathcal{T}|+|\mathcal{P}|\ll |\mathcal{R}|\times|\mathcal{T}|\times |\mathcal{P}|=N_G$. 
Accordingly, the complexity (per iteration) is now $\mathcal{O}(M (|\mathcal{R}|+|\mathcal{T}|+|\mathcal{P}| + \lvert \mathcal{V} \rvert))$, which is much lower than that of the initial Algorithm~\ref{alg:algo1}.
\begin{algorithm}[t]
\caption{Reduced-Complexity \ac{RIS} Configuration Design}
\label{alg_lowcomp}
\textbf{Initialize:} $\beta\in (0,1)$, $\omegab^{(0)}=\text{proj}_{\mathcal{V}}( \sum_{\text{p} \in \{ \rho,\theta,\phi\}}\BB_{\text{p}}^\dagger \gg_{\text{p}})$.
\begin{algorithmic}[1]
\For {$r=1,2,\ldots$} 
\State Update the scaling factor as:\begin{align}
    s^{(r)}= \sum_{\text{p} \in \{ \rho,\theta,\phi\}}\frac{(\omegab^{(r-1)})^{\h}\BB_{\text{p}}^{\h}\gg_{\text{p}}}{\Vert\BB_{\text{p}}\omegab^{(r-1)} \Vert^2}.
    \end{align}
\State Update the \ac{RIS} configuration as:
\begin{align}
    & \omegab^{(r)}_{\text{u}}=\omegab^{(r-1)}+\\
    & \beta \sum_{\text{p} \in \{ \rho,\theta,\phi\}}\frac{
    (s^{(r)})^{*}\BB_{\text{p}}^{\h}(\gg_{\text{p}}-s^{(r)}\BB_{\text{p}}\omegab^{(r-1)})}{\lambda_{\text{max}}(|s^{(r)}|^{2}\BB_{\text{p}}^{\h}\BB_{\text{p}})}\notag 
\end{align}
\State Perform the projection: $\omegab^{(r)}=\text{proj}_{\mathcal{V}}(\omegab^{(r)}_{\text{u}})$.
\EndFor
\end{algorithmic}
\end{algorithm}
\section*{Application to Location Oriented RIS Beams Design}


\subsection*{FIM and PEB}
In our specific localization application context, we first define the vector of \ac{UE} position and channel parameters in the 3D spherical coordinates system, as $\upzetab_\text{sph}=[ \rho, \theta, \phi,  \alpha_{r}, \alpha_{i}]^\top \in \mathbb{R}^{5\times1}$, and compute the \acs{FIM} accordingly \cite[Chapter~3.7]{kay_fundamentals}.
 \begin{align}
     \JJ_{\text{sph}}(\upzetab_{\text{sph}}) =
    \frac{2E_s}{N_0}  \text{Re} \left\{
    \left(\frac{\partial \upmub} {\partial\upzetab_\text{sph}}\right)^{\mathsf{H}} \frac{\partial \upmub} {\partial\upzetab_{\text{sph}}}
   \right\} 
    \in \mathbb{R}^{5\times 5}, \label{eq:FIMsph}
 \end{align}
where $\upmub = \alpha \FF^{\top} \aa(\pp_{\text{UE}})$ refers to the noiseless part of the observation and 
 %
 \begin{align}
 \left[\frac{\partial \upmub}{\partial {\rho}},\frac{\partial \upmub}{\partial {\theta}},  \frac{\partial \upmub}{\partial 
 {\phi}} \right]& =\alpha \FF^\top\left[ \Dot{\aa}_{\rho}(\pp_{\text{UE}}), \Dot{\aa}_{\theta}(\pp_{\text{UE}}), \Dot{\aa}_{\phi}(\pp_{\text{UE}})\right]\\
     \left[\frac{\partial \upmub}
    {\partial {{\alpha}}_{r}},\frac{\partial \upmub}
    {\partial {{\alpha}}_{i}}\right] &= \FF^\top \aa(\pp_{\text{UE}}) [1,\jmath], 
\end{align}
with $\Dot{\aa}_{x}(\pp_{\text{UE}}) = \nicefrac{\partial \aa(\pp_{\text{UE}})}
{\partial x}\in \mathbb{C}^{M\times 1}$.

After introducing the corresponding set of parameters in the Cartesian coordinates system, that is, $\upzetab_\text{car}=[ \pp_{\text{UE}}^{\top}, \alpha_{r}, \alpha_{i}]^\top \in \mathbb{R}^{5\times1}$ with $\pp_{\text{UE}}^{\top}=[x_{\text{UE}},y_{\text{UE}},z_{\text{UE}}]^{\top}$, the Jacobian $\CC = \nicefrac {\partial\upzetab_{\text{sph}} } {\partial\upzetab_{\text{car}}}$ is used to re-compute the previous \acs{FIM} as
\begin{align}\label{eq:carFIM}
    & \JJ_\text{car}(\upzetab_\text{car}) = \CC^{\top} \JJ_\text{sph}(\upzetab_\text{sph}) \CC.
\end{align}
Computing the \ac{FIM} in the spherical domain followed by a transformation to the cartesian domain is a better approach compared to directly calculating it in the latter. This is because the derivative beams used in the computation require access to the spherical positional variables.
Finally, the best achievable positioning performance is characterized by means of the following \ac{PEB}, which represents a lower bound on the accuracy of any unbiased location estimator \cite[Chapter~2.4.2]{vanTrees}
\begin{align}
\mathrm{PEB}(\FF;\upzetab_\text{car}) & = \sqrt{\mathrm{tr}\left(\left[\JJ_\text{car}^{-1}(\upzetab_\text{car})\right] _{(1:3,1:3)} \right)}\label{eq:PEB}\\
& \le \sqrt{\mathbb{E}\left\{ \Vert\pp_{\text{UE}}-\hat{\pp}_{\text{UE}} \Vert^2\right\}},
\end{align}
where $\hat{\pp}_{\text{UE}}$ stands for any unbiased estimate of $\pp_{\text{UE}}$ and we have made the dependence on the precoding matrix $\FF$ explicit.

In the following, the use of this \ac{PEB} is twofold. First, it is exploited as a parametric optimization objective to determine a localization-optimal \ac{RIS} configuration suited to the stated downlink \ac{NF} \ac{NLoS} positioning problem (See the next section). Beyond, the \ac{PEB} will be used also as a general performance indicator to assess and benchmark the impact of beam approximation on localization for several \ac{RIS} configurations and different \ac{RIS} lookup tables (See the numerical results section).

\subsection*{PEB Minimization}
Assuming the \ac{UE} position to be known a priori, the power-constrained \ac{PEB} optimization problem, as a function of the \ac{RIS} configuration, is first formulated as follows
\vspace{-3mm}
\begin{subequations}
\begin{align} \label{eq:PEB_optimization}
     \min_{\FF} \quad &  \mathrm{PEB}(\FF;\upzetab_\text{car}) \\
     \text{s.t.} \quad & \mathrm{tr}(\FF\FF^{\mathsf{H}})=MT. 
\end{align}
\end{subequations}
Similar to~\cite{Rahal_Localization-Optimal_RIS}, using the change of variable $\XX=\FF\FF^{\mathsf{H}}$ and removing the constraint $\text{rank}(\XX)=T$ \cite[Chapter~7.5.2]{boyd2004convex} \cite{garcia_optimal_2018}, it can be shown
%
%
from~\cite[Appendix~C]{4359542}  
that the optimal 
matrix solution $\XX^{\smallstar}$ to the equivalent relaxed convex \ac{SDP} problem is of the specific form 
\begin{align}
    \XX^{\smallstar} = \UU \Lambdab \UU^{\mathsf{H}} 
    \label{eq:Xopt}
\end{align}
where $\Lambdab \in \mathbb{C}^{4\times4}$ is a \ac{PSD} matrix with the beam weights to be applied onto -or equivalently, the relative powers allocated to- the columns of $\UU$ lying on its diagonal.
\begin{align}
    \UU \triangleq [
\aa^*(\pp_{\text{UE}}) \hspace{1mm}
\Dot{\aa}^*_{\rho}(\pp_{\text{UE}}) \hspace{1mm}
\Dot{\aa}^*_{\theta}(\pp_{\text{UE}}) \hspace{1mm}
\Dot{\aa}^*_{\phi}(\pp_{\text{UE}})
]. \label{eq:defU}
\end{align}
The columns of $\UU$ can be physically interpreted as the \ac{RIS} steering vector and its successive derivatives with respect to the spherical coordinates, similar to that involved in \eqref{eq:FIMsph}.
Note that the space spanned by these columns could also be spanned by 4 orthonormalized vectors, after application of the Gram-Schmidt algorithm, so that $\UU^{\mathsf{H}} \UU=M\II_4$. 
This step enables us to re-write the initial optimization constraint $\mathrm{tr}(\XX)=MT$ as $\mathrm{tr}(\Lambdab)=T$. 
After performing the transformation above (i.e., from $\XX \in \mathbb{C}^{M\times M}$ to $\Lambdab \in \mathbb{C}^{4\times 4}$), and by applying schur's compliment on~\ref{eq:PEB_optimization} to eliminate the inverse of the information matrix term, an equivalent low-complexity optimization problem can be simply stated as follows 
\begin{subequations} \label{finalOpt}
\begin{align} 
    \min_{\Lambda\uu} \quad & \boldone^{\top}\uu \\
     \text{s.t.} \quad 
     & 
     \begin{bmatrix}
    \JJ_\text{car} & \ee_{k} \\ 
    \ee_{k}^{\top} & u_k 
    \end{bmatrix} \succeq 0, k = 1, 2, 3 ,  \\
     & \mathrm{tr}(\Lambdab)=T, \\
     & \Lambdab \succeq 0.
\end{align}
\end{subequations}
Furthermore, the problem can be further relaxed by forcing $\Lambdab$ to be diagonal, i.e., $\Lambdab=\text{diag}(\lambdab)$. The entries of vector $\lambdab$ can hence be viewed as relative power allocations (or equivalently, as relative transmission periods within a time-sharing approach, depending on the implementation \cite{Rahal_Localization-Optimal_RIS}), which must be assigned to the different columns of $\UU$. In the following, we consider solving the final optimization problem \eqref{finalOpt} through CVX \cite{cvx}.


\subsubsection*{Localization-optimal RIS Beams Approximation}

From \eqref{eq:defU}, if $\pp_{\text{des}} = \pp_{\text{UE}}$ denotes the position of one single \ac{UE} to be localized (i.e., $I=1$), it comes that 4 distinct desired beams must be synthesized in practice so as to optimize localization performance, as follows:
\begin{itemize}
\item \textbf{1 directional beam:} $G(\pp)\propto  (\bb^*(\pp_{\text{des}},\pptx))^\top \bb(\pp,\pptx)$, where $\propto$ indicates proportionality (to avoid normalization issues). Here, $\bb(\pp_{\text{des}},\pptx) = \aa(\pp_{\text{des}}) \odot \aa(\pptx) $, with $\aa(\pp_{\text{des}})$ representing the desired directional beam pointing towards the UE position.

%
\item \textbf{3 derivative beams}: $G_x(\pp)\propto  (\dot{\bb}_{x}^*(\pp_{\text{des}},\pptx))^\top \bb(\pp,\pptx)$, where 
$\dot{\bb}_{x}(\pp_{\text{des}},\pptx) = \partial {\bb}(\pp_{\text{des}},\pptx) / \partial x$, in which $x$ represents alternatively the range $\rho$, elevation angle $\theta$, and azimuth angle $\phi$ (see Fig.~\ref{fig:Geometry}). Here, $\dot{\bb}_{x}(\pp_{\text{des}},\pptx) = \dot{\aa}_{x}(\pp_{\text{des}}) \odot \aa(\pptx) $, with $\dot{\aa}_{x}(\pp_{\text{des}})$ representing the desired derivative beam pointing towards the UE position.
%
\end{itemize}

\section*{Numerical Results}


\subsection*{Simulation Parameters}
\begin{figure}[ht]
 \centering
 \resizebox{0.8\columnwidth}{!}{
%
%
\begin{tikzpicture}

\begin{axis}[%
width=4.521in,
height=4.521in,
at={(0.758in,0.481in)},
scale only axis,
xmin=-1.1,
xmax=1.1,
ymin=-1.1,
ymax=1.1,
axis x line*=bottom,
axis y line*=left,
xmajorgrids,
ymajorgrids,
xlabel = Real,
ylabel = Imagenary,
xlabel style = {font = \normalsize},
ylabel style = {font = \normalsize},
axis background/.style={fill=white},
title style={font=\bfseries},
legend style={font = \large, legend cell align=left, align=left, draw=white!15!black}
]
%
\addplot [color=black, line width=1.2pt, forget plot]
  table[row sep=crcr]{%
1	0\\
0.997986676471884	0.0634239196565645\\
0.991954812830795	0.126592453573749\\
0.981928697262707	0.18925124436041\\
0.967948701396356	0.251147987181079\\
0.950071117740945	0.312033445698487\\
0.928367933016073	0.371662455660328\\
0.902926538286621	0.429794912089172\\
0.873849377069785	0.486196736100469\\
0.841253532831181	0.540640817455598\\
0.805270257531059	0.59290792905464\\
0.766044443118978	0.642787609686539\\
0.72373403810507	0.690079011482112\\
0.678509411557132	0.734591708657533\\
0.630552667084523	0.776146464291757\\
0.580056909571198	0.814575952050336\\
0.527225467610502	0.849725429949514\\
0.472271074772683	0.881453363447582\\
0.415415013001886	0.909631995354518\\
0.356886221591872	0.934147860265107\\
0.296920375328275	0.954902241444074\\
0.235758935509427	0.971811568323542\\
0.173648177666931	0.984807753012208\\
0.110838199901011	0.993838464461254\\
0.0475819158237424	0.998867339183008\\
-0.015865963834808	0.999874127673875\\
-0.0792499568567885	0.996854775951942\\
-0.142314838273285	0.989821441880933\\
-0.204806668065191	0.978802446214779\\
-0.266473813690035	0.963842158559942\\
-0.327067963317421	0.945000818714669\\
-0.386345125693128	0.922354294104581\\
-0.444066612605774	0.895993774291336\\
-0.5	0.866025403784439\\
-0.55392006386611	0.832569854634771\\
-0.605609687137666	0.795761840530832\\
-0.654860733945285	0.755749574354258\\
-0.701474887706321	0.712694171378863\\
-0.745264449675755	0.666769000516292\\
-0.786053094742787	0.618158986220605\\
-0.823676581429833	0.567059863862771\\
-0.857983413234977	0.513677391573407\\
-0.888835448654923	0.458226521727411\\
-0.916108457432069	0.400930535406614\\
-0.939692620785908	0.342020143325669\\
-0.959492973614497	0.28173255684143\\
-0.975429786885407	0.220310532786541\\
-0.987438888676394	0.15800139597335\\
-0.995471922573085	0.0950560433041829\\
-0.999496542383185	0.0317279334980681\\
-0.999496542383185	-0.0317279334980679\\
-0.995471922573085	-0.0950560433041826\\
-0.987438888676394	-0.15800139597335\\
-0.975429786885407	-0.220310532786541\\
-0.959492973614497	-0.281732556841429\\
-0.939692620785908	-0.342020143325669\\
-0.91610845743207	-0.400930535406613\\
-0.888835448654924	-0.45822652172741\\
-0.857983413234977	-0.513677391573406\\
-0.823676581429833	-0.567059863862771\\
-0.786053094742788	-0.618158986220605\\
-0.745264449675755	-0.666769000516292\\
-0.701474887706322	-0.712694171378863\\
-0.654860733945285	-0.755749574354258\\
-0.605609687137667	-0.795761840530832\\
-0.55392006386611	-0.832569854634771\\
-0.5	-0.866025403784438\\
-0.444066612605774	-0.895993774291336\\
-0.386345125693129	-0.922354294104581\\
-0.327067963317422	-0.945000818714668\\
-0.266473813690035	-0.963842158559942\\
-0.204806668065191	-0.978802446214779\\
-0.142314838273285	-0.989821441880933\\
-0.0792499568567888	-0.996854775951942\\
-0.0158659638348076	-0.999874127673875\\
0.0475819158237424	-0.998867339183008\\
0.110838199901011	-0.993838464461254\\
0.17364817766693	-0.984807753012208\\
0.235758935509427	-0.971811568323542\\
0.296920375328275	-0.954902241444074\\
0.356886221591872	-0.934147860265107\\
0.415415013001886	-0.909631995354519\\
0.472271074772682	-0.881453363447582\\
0.527225467610502	-0.849725429949514\\
0.580056909571198	-0.814575952050336\\
0.630552667084522	-0.776146464291757\\
0.678509411557132	-0.734591708657534\\
0.723734038105069	-0.690079011482113\\
0.766044443118977	-0.64278760968654\\
0.805270257531059	-0.59290792905464\\
0.841253532831181	-0.540640817455597\\
0.873849377069785	-0.486196736100469\\
0.902926538286621	-0.429794912089172\\
0.928367933016072	-0.371662455660328\\
0.950071117740945	-0.312033445698487\\
0.967948701396356	-0.251147987181079\\
0.981928697262707	-0.189251244360411\\
0.991954812830795	-0.12659245357375\\
0.997986676471884	-0.0634239196565654\\
1	-2.44929359829471e-16\\
};
\addplot [color=white!80!black, line width=1.2pt, forget plot]
  table[row sep=crcr]{%
-1	0\\
1	0\\
};
\addplot [color=white!80!black, line width=1.2pt, forget plot]
  table[row sep=crcr]{%
0	-1\\
0	1\\
};
\addplot [color=gray, only marks, mark=asterisk, mark options={solid, mark size=4pt, thick, gray}]
  table[row sep=crcr]{%
0.705881663503301	0.454874816836335\\
0.614312562634405	0.531271558748285\\
0.475934107814557	0.506942448025792\\
0.312209976375841	0.42259048294233\\
0.111822453204272	0.312651838583432\\
-0.0913794808614622	-0.0208313626559627\\
0.135183293840408	-0.446930997789427\\
0.457051113855161	-0.537546263774266\\
0.646418002339778	-0.46806616159338\\
0.830757906262032	-0.357145071975791\\
0.881458766139799	-0.254203355481816\\
0.924391111031014	-0.207693362330688\\
0.926939342567385	-0.162193617949656\\
0.943816578371453	-0.114314640581941\\
};
\addlegendentry{$\mathcal{V}$}


\addplot [color=green, only marks,  mark=o, mark options={solid, mark size=4pt, thick, green}]
  table[row sep=crcr]{%
0.891250938133746	0\\
-0.891250938133746	1.11022302462516e-16\\
};
\addlegendentry{$\mathcal{K}$1}

\addplot [color=blue, only marks, mark=+, mark options={solid, mark size=5pt, thick, blue}]
  table[row sep=crcr]{%
0.891250938133746	0\\
0	0.891250938133746\\
-0.891250938133746	1.11022302462516e-16\\
-1.11022302462516e-16	-0.891250938133746\\
};
\addlegendentry{$\mathcal{K}$2}
\node[align=center, font = \large]
at (axis cs:0.1,0.05) {(0,0)};
\node[align=center, font = \LARGE]
at (axis cs:0,0) {.};
\end{axis}

\begin{axis}[%
width=5.833in,
height=4.375in,
at={(0in,0in)},
scale only axis,
xmin=0,
xmax=1,
ymin=0,
ymax=1,
axis line style={draw=none},
ticks=none,
axis x line*=bottom,
axis y line*=left
]
\end{axis}
\end{tikzpicture}%
}
 \caption{The sets $\mathcal{V}$, $\mathcal{K}1$, and $\mathcal{K}2$ with the values for the \ac{RIS} elements responses, as well as the unit-modulus set, plotted in the complex plane. 
 }
 \label{fig:set_distribution}
 \end{figure}
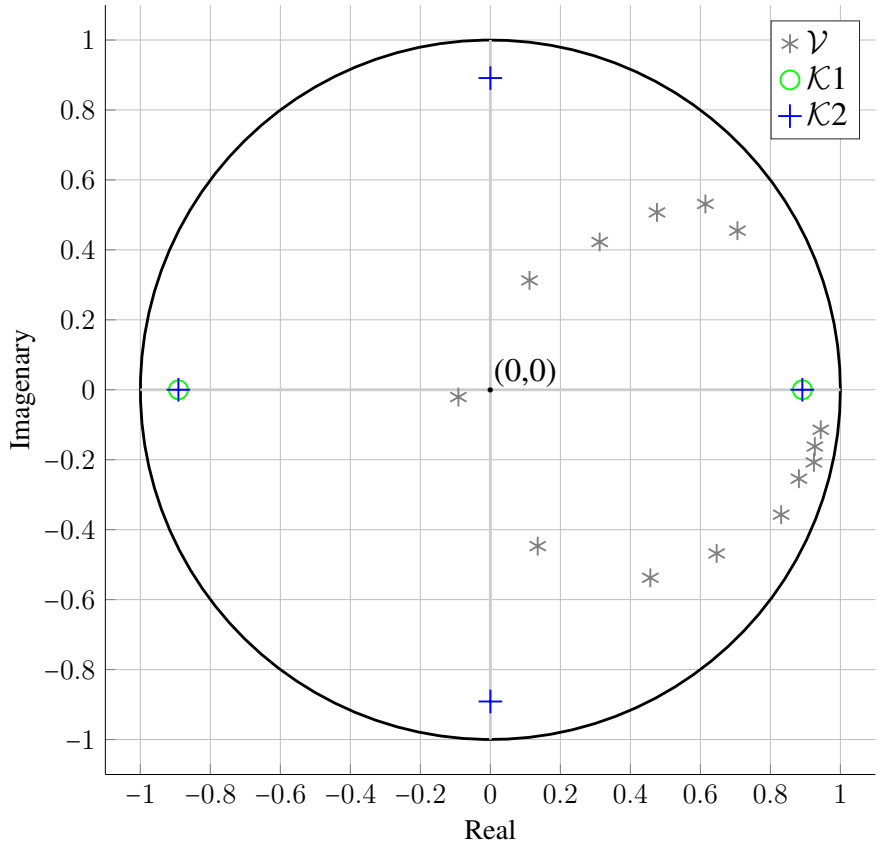

To assess the performance of the proposed beam synthesis method, we considered actual lookup tables accounting for the \ac{RIS} response per unit element of distinct hardware prototypes, i.e., distinct sets of feasible complex reflection coefficients per element. These prototypes have been developed and characterized in the frame of the H2020 {RISE-6G}\footnote{See \url{https://RISE-6G.eu} for more information.} project. In particular, we have considered the following sets (See Fig.~\ref{fig:set_distribution}):
\begin{itemize}   
    \item $\mathcal{V}$ \cite[Table 1]{fara2021prototype}: a first set characterizing a prototype based on a single-diode varactor, with $14$ possible complex values per \ac{RIS} element;
    \item $\mathcal{K}1$ \cite{DiPalma_2017}: a second set characterizing a prototype revisiting a transmit-array architecture \cite{DiPalma_2017} based on p-i-n diodes and enabling $1$-bit phase quantization, hence with $2$ possible complex values per \ac{RIS} element;
    \item $\mathcal{K}2$: a third set similar to $\mathcal{K}1$, but with $2$-bit phase quantization and hence, 4 possible complex values per \ac{RIS} element.
\end{itemize}
The performance obtained with each of the previous sets after beam approximation is compared with that of an idealized variant (so-called \emph{unconstrained}), where the \ac{RIS} complex element-wise reflection coefficients all lie on the unit circle with continuous phase values.

In our simulations, the carrier frequency was set at $5.15\,\mathrm{GHz}$ for $\mathcal{V}$ and $28\,\mathrm{GHz}$ for the other sets, 
reflecting the actual operating frequency of each RIS hardware prototype. And to make sure that the localization performance comparison is fair, we ensured that the propagation loss of the \ac{BS}-\ac{RIS} \ac{RIS}-\ac{UE} paths is equal disregarding the operating frequency effect.

In terms of addressed scenario, without loss of generality, we focus on a canonical configuration, where $\pptx$ is set to $[3, 3, 0]^\top~\mathrm{m}$ and the RIS is placed at the origin, i.e., $\pp_{\text{RIS}}=[0, 0, 0]^\top~\mathrm{m}$. 
For beam patterns visualization and beam fidelity analysis (See Fig.~\ref{fig:directional} to Fig.~\ref{fig:2D-K2}), we first consider a desired point $\pp_{\text{des}}\equiv \pp_{\text{UE}}$ located in $[0, 2, 0]^\top~\mathrm{m}$ for both steering and derivative beam patterns. The 1D visualizations are intended to show the performance in terms of maximum power gain and width, allowing us to evaluate the designed beams in the desired direction, whereas the 2D illustrations showcase the beam pattern in terms of both the azimuth and elevation angles, hence focusing on the overall angular behavior, as well as on the potential presence of harmful grating lobes. These visualizations include the magnitude $|G(\boldsymbol{p})|$ in 1D or 2D slices in spherical dimensions.
%
In Fig.~\ref{fig:PEBdist}, the \ac{PEB} is evaluated as a function of the \ac{RIS}-\ac{UE} distance and accordingly, the \ac{UE} position is set to $[-r,~r,~r]^{\top}~\mathrm{m}$ where $r$ varies between $0.2~\mathrm{m}$ and $10~\mathrm{m}$. However, in Fig.~\ref{fig:PEBaz}, we set the spherical coordinates of the user in a different way where the \ac{RIS}-\ac{UE} distance is set to a constant value $\rho = 2~\mathrm{m}$, the elevation angle is also set to a constant $\phi = \frac{\pi}{2}~\mathrm{rad}$; whereas $\theta$, the azimuth component, varies across the entire defined range, i.e., $[0~\cdots~\pi]~\mathrm{rad}$ (excluding the first and last values where the \ac{UE} is co-planar with the \ac{RIS}). 
Similarly, in Fig.~\ref{fig:PEBel}, the range is set to $\rho = 2~\mathrm{m}$, $\theta = \frac{\pi}{2}~\mathrm{rad}$ and now the elevation angle $\phi$ varies across the $[0~\cdots~\pi]~\mathrm{rad}$ range. 
For this positioning performance assessment, we compare the \ac{PEB} when applying ideal localization-optimal RIS beams with those obtained via the  proposed approximation method. For benchmark purposes, similar \ac{PEB} evaluations are also carried out with random and directional RIS configurations where we adopted a monte-carlo approach and we averaged the \ac{PEB} over all the random trials. 
In the former case, the complex \ac{RIS} element-wise reflection coefficients are simply drawn randomly from the corresponding set for each prototype, with no further projection.

 The main simulation parameters (system model and scenario) are summarized in Table~\ref{table:params}.
\begin{table}
\centering
\caption{General simulation parameters.}
\resizebox{0.8\columnwidth}{!} {
\begin{tabular}{ |c|c||c|c| } 
 \hline
 parameter & value & parameter & value\\
 \hline
 $f_{c}$ & $5.15-28~\mathrm{GHz}$ & $E_{\text{tot}}$ & $(\nicefrac{P_{\text{tx}}}{W})MT$\\
 $W$ & $120~\mathrm{kHz}$ & $P_{\text{tx}}$ & $20~\mathrm{dBm}$\\ 
 noise figure $n_f$ & $8~\mathrm{dB}$ & RIS size & $M = 32\times 32$ elements\\
   $N_0$ & $-174~\mathrm{dBm/Hz}$ & transmissions & $T = 40$\\
 \hline
\end{tabular}\vspace{-4mm}
}
\label{table:params}
\vspace{-4mm}
\end{table}

\subsection*{Results and Discussion}
\subsubsection*{Qualitative Analysis of Approximated RIS Beams}

\begin{figure}[ht]
 \centering
 \resizebox{0.9\columnwidth}{!}{\input{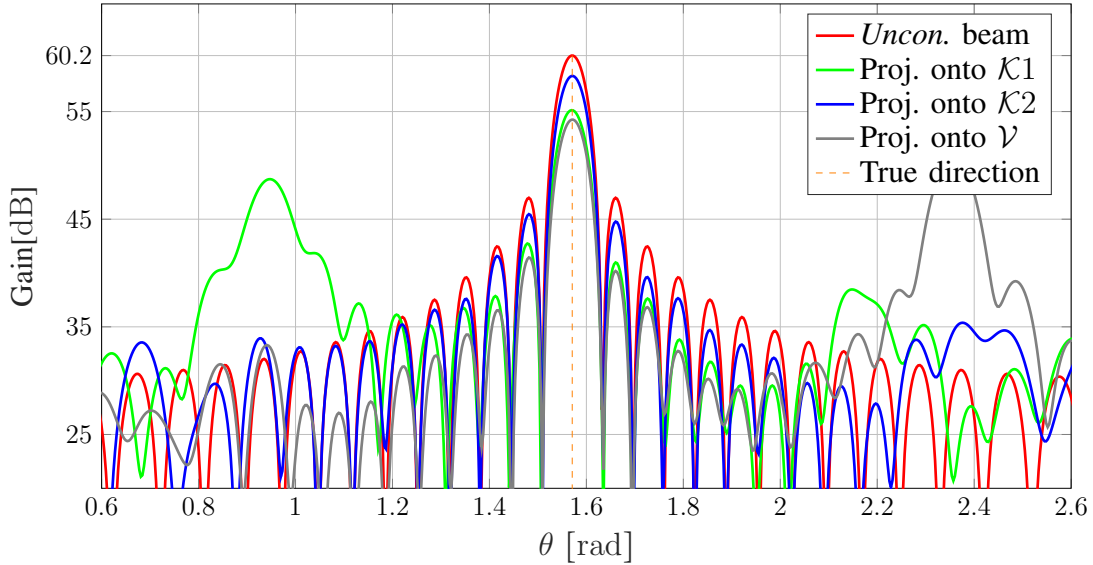}}
 \caption{Steering beam patterns as a function of the azimuth angle $\theta$ for various beam synthesis methods under gradual \ac{RIS} hardware constraints, including the realistic \ac{RIS} element responses of \cite{DiPalma_2017,fara2021prototype}.}
 \label{fig:directional}
 \end{figure}

\paragraph*{1D Visualization}
In Fig.~\ref{fig:directional} and \ref{fig:derivative}, we show approximated patterns for both 1D steering and derivative beams, as a function of the azimuth parameter $\theta$, while maintaining the other coordinates at their true (i.e., desired) values. These patterns have been generated without imposing any constraint on the RIS precoder (\emph{unconstrained}) or by utilizing realistic hardware limitations derived from the lookup tables (sets $\mathcal{K}1$, $\mathcal{K}2$, and $\mathcal{V}$). It is important to note that the projection onto any set can be achieved through a single step or can be improved through iterative refinement. The latter approach have been adopted to achieve a higher level of precision.

In Fig.~\ref{fig:directional}, we notice at first that the main lobe of the \emph{unconstrained} steering beam (red curve) has a peak of $60.2 \mathrm{dB}$, which is in line with the beamforming gain offered by the considered \ac{RIS} of $M=32\times32$ elements. 
Moreover, comparing the beam projected onto the set $\mathcal{K}1$ (i.e., with $1$-bit unit cells; green curve) with the \emph{unconstrained} one, we see a significant loss, which can be mitigated by about $3 \mathrm{dB}$, by projecting the beam onto $\mathcal{K}2$ instead ($2$-bit unit cells; blue curve).
Furthermore, projecting the beam onto set $\mathcal{V}$, results in, as expected, more degradation in the beam peak value.
Beyond, if we take a look at the entire beam shape across $\theta$, we notice that constraining into real sets leaves the beam with unwanted secondary lobes, explaining also the power loss experienced at the peak of the main lobe. Table \ref{table:peakValues} summarizes the main indicators regarding beam fidelity issues, in terms of power loss in the desired direction and the presence of unwanted secondary lobes aside. It simply indicates the peak power value of the main lobe as well as the position and the peak value of the secondary lobes, across different prototypes, which are defined by the highest peaks breaking the \emph{unconstrained} envelope.
\begin{table}[ht]
\centering
\caption{Peak Gain Distribution of Main and Secondary Lobes Across Different \ac{RIS} Prototypes}
\resizebox{0.8\columnwidth}{!}{
  \begin{tabular}{c|c|c|c|c}
    & \emph{Unconstrained} & $\mathcal{K}1$ & $\mathcal{K}2$ & $\mathcal{V}$\\ \hline
    Main Lobe Peak Value~[$\mathrm{dB}$] & 60.2 & 55.2 & 58.2 & 54.2\\ 
    Secondary Lobe Peak Value~[$\mathrm{dB}$] & - & 48.5  & 35.36 & 50.7 \\ 
    Position of Secondary Lobe $\theta~[\mathrm{rad}]$ & - & 0.9 & 2.37 & 2.3\\
  \end{tabular}
}
  \label{table:peakValues}
\end{table}

Fig.~\ref{fig:derivative} shows that the generated derivative beam patterns exhibit very similar trends to that observed with the steering ones. 
Additionally, as the figure displays the derivative with respect to the azimuth angle, a null is present when $\theta$ corresponds to the desired direction, as already pointed out in~\cite{keskin_optimal_2021}. 

 \begin{figure}[ht]
 \centering
 \resizebox{0.9\columnwidth}{!}{\input{figures/beams1D_derivative.tex}}
 \caption{Derivative beam patterns as a function of the azimuth angle $\theta$ for various beam synthesis methods under gradual \ac{RIS} hardware constraints, including the realistic \ac{RIS} element responses of \cite{DiPalma_2017,fara2021prototype}.}
 \label{fig:derivative}
 \end{figure}
\paragraph*{2D Visualization}

\begin{figure}[ht]
    \centering
    \begin{subfigure}[b]{0.48\textwidth}
        \centering
        \includegraphics[width=\textwidth]{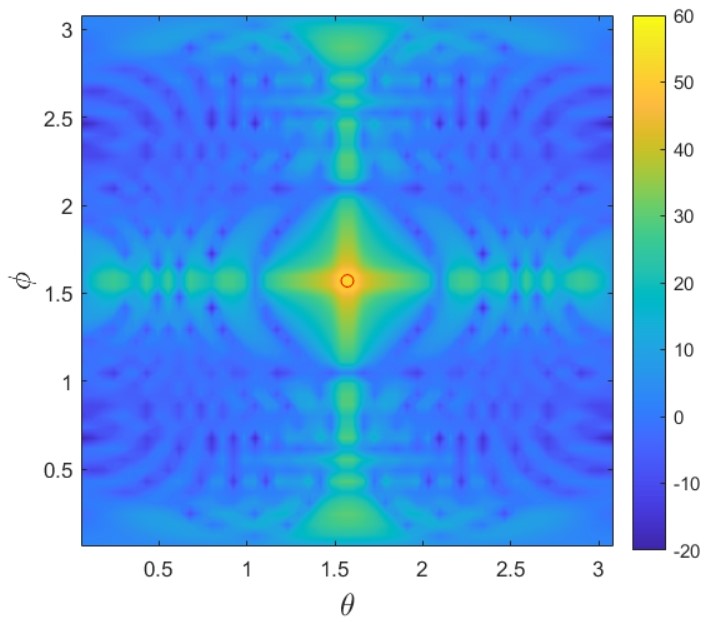}
        \caption[]%
        {Example of the \emph{unconstrained} steering RIS beam visualization in 2D.}    
        \label{fig:2D-desired}
    \end{subfigure}
        \hfill
    \begin{subfigure}[b]{0.48\textwidth}  
            \centering 
        \includegraphics[width=\textwidth]{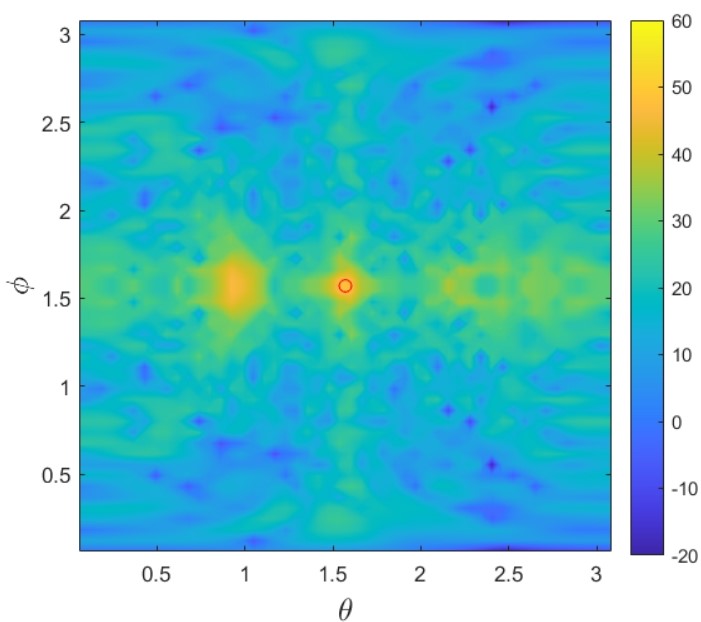}
        \caption[]%
        {Designed RIS beam resulting from the projection to $\mathcal{K}1$ (Same example as in Fig.~\ref{fig:2D-desired}).}    
        \label{fig:2D-K1}
    \end{subfigure}
    \vskip\baselineskip
    \begin{subfigure}[b]{0.48\textwidth}  
        \centering 
        \includegraphics[width=\textwidth]{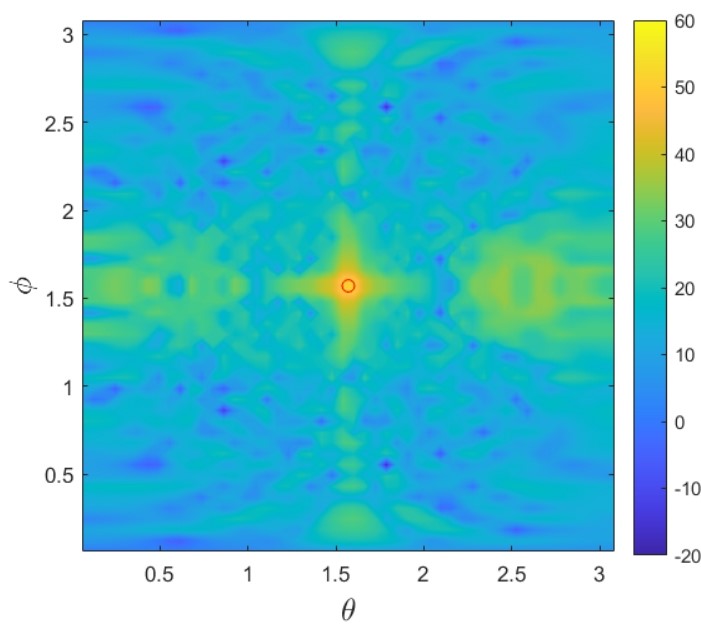}
        \caption[]%
        {Designed RIS beam resulting from the projection to $\mathcal{K}2$ (Same example as in Fig.~\ref{fig:2D-desired}).}    
        \label{fig:2D-K2}
    \end{subfigure}
   \hfill
    \begin{subfigure}[b]{0.48\textwidth}  
        \centering 
        \includegraphics[width=\textwidth]{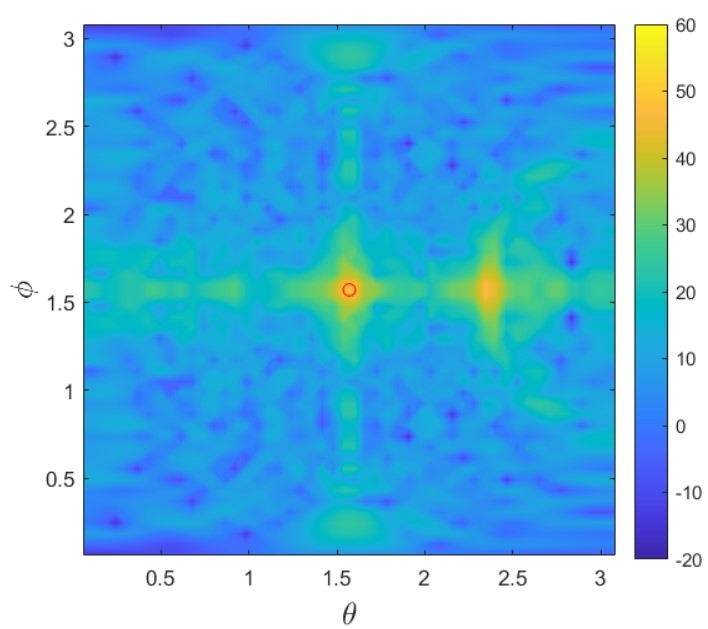}
        \caption[]%
        {Designed RIS beam resulting from the projection to $\mathcal{V}$ (Same example as in Fig.~\ref{fig:2D-desired}).}    
        \label{fig:2D-V}
    \end{subfigure}
    \caption[]
    {Illustration of the steering beam pattern pointing to one single desired direction (red circle), as a function of angles $\phi$ (elevation) and $\theta$ (azimuth).} 
        \label{fig:2D-beams}
\end{figure}

The heatmaps in Fig.~\ref{fig:2D-desired} to Fig.~\ref{fig:2D-V} display, as a function of the direction of departure from the \ac{RIS} (i.e., of both azimuth and elevation angles) and a specified desired direction (red circle), the \emph{unconstrained} steering beam, as well as its projections onto sets $\mathcal{K}1$, $\mathcal{K}2$, and $\mathcal{V}$, respectively. In Fig.~\ref{fig:2D-desired}, we can see that only one main lobe is present in the desired direction, as expected. Note that this is used as a reference for comparing the beams constrained with realistic \ac{RIS} hardware.

Fig.~\ref{fig:2D-K1}, visualizes the beam projection onto set $\mathcal{K}1$, where the main lobe is still observed in the desired direction, even though clearly attenuated in comparison with the \emph{unconstrained} beam from Fig.~\ref{fig:2D-desired}. The presence of a strong and systematic secondary grating lobe is also noted, which turns out to be a standard reflection, regardless of the desired beam direction, number of \ac{RIS} elements, or inter-elements spacing. This kind of grating lobe  arises due to the severe quantization of the \ac{RIS} element phase, 
creating some kind of spatial aliasing. 
However, as shown on Fig.~\ref{fig:2D-K2}, this problem of grating lobe can be mitigated after adding only one more bit of phase quantization. In this case, a higher peak value (by about $+3~\mathrm{dB}$) is also achieved for the main lobe in the desired direction, even if the levels of all the other secondary lobes remain globally high and comparable to those in the $1$-bit phase quantization case.
Lastly, Fig.~\ref{fig:2D-V} depicts the beam pattern gain resulting from the projection onto $\mathcal{V}$. Regardless of the high number of quantization levels (i.e., $14$) shown in Fig.~\ref{fig:set_distribution}, a secondary lobe is also sill present pointing towards a fixed unwanted direction, which corresponds to a natural specular reflection. This grating lobe comes from the fact that set $\mathcal{V}$ is not centered at the origin of the complex plane (See Fig.~\ref{fig:set_distribution}), but shifted and significantly down-scaled in comparison with the unit circle, due to amplitude losses. Accordingly, the $2\pi~\mathrm{rad}$ phase domain is not entirely covered by the feasible reflection coefficients per element. Indeed, given the span of valid phases observed in $\mathcal{V}$ (say, reflection coefficients experiencing attenuations lower that $5~\mathrm{dB}$), which covers less than $\pi~\mathrm{rad}$ in practice (i.e., less than what would be feasible with a $1$-bit quantization of the \ac{RIS} element phase in $\mathcal{K}1$), this unwanted reflection was hence expected with $\mathcal{V}$ too.
In addition to the main grating lobe, the intensity of all the secondary lobes is also generally higher throughout the entire 2D area. 

\subsubsection*{Quantitative Analysis of Positioning Performance with Approximated RIS Beams}
\begin{figure}[ht]
     \centering
     \resizebox{\columnwidth}{!}{
%
\begin{tikzpicture}

\begin{axis}[%
width=4.5in,
height=3in,
at={(0.758in,0.481in)},
scale only axis,
xmin=1.4,
xmax=17.5,
xlabel style={font=\large \color{white!15!black}},
xlabel={RIS-UE distance $[\mathrm{m}]$},
ymin=0.001,
ymax=10,
ylabel style={font=\large \color{white!15!black}},
ylabel={PEB $[\mathrm{m}]$},
yminorticks=true,
ymode = log,
axis background/.style={fill=white},
xmajorgrids,
yminorgrids,
legend style={font=\footnotesize, at={(0.65,0.03)}, anchor=south west, legend cell align=left, align=left, draw=white!15!black}
]
\addlegendimage{empty legend}
\addlegendentry{\hspace{-.6cm}\textbf{Optimal Design}}
\addplot [color=red, line width=1.2pt, mark=square]
  table[row sep=crcr]{%
0.3464	  0.0000127\\
0.8660	  0.00019\\
1.7320	  0.00126\\
3.4641	  0.00552\\
5.1961	  0.01796\\
6.9282	  0.04316\\
8.6602	  0.08721\\
10.392	  0.15751\\
13.856	  0.41196\\
17.320	  0.70435\\
};
\addlegendentry{\emph{Uncon.} beam}

\addplot [color=green, line width=1.2pt, mark=o, mark size = 3]
  table[row sep=crcr]{%
0.3464            0.00016\\
0.8660            0.00064\\
1.7320            0.00401\\
3.4641            0.02040\\
5.1961            0.07211\\
6.9282            0.18139\\
8.6602            0.35628\\
10.392            0.54593\\
13.856            1.18583\\
17.320            2.32581\\
};
\addlegendentry{Proj. onto $\mathcal{K}1$}
\addplot [color=blue, line width=1.2pt, mark=+, mark size = 3]
  table[row sep=crcr]{%
0.3464             0.00012\\
0.8660             0.00044\\
1.7320             0.00417\\
3.4641             0.01362\\
5.1961             0.04865\\
6.9282             0.12531\\
8.6602             0.27047\\
10.392             0.41203\\
13.856             0.90976\\
17.320             1.62787\\
};
\addlegendentry{Proj. onto $\mathcal{K}2$}
\addplot [color=gray, line width=1.2pt, mark=asterisk, mark size = 3]
  table[row sep=crcr]{%
0.3464          0.00016\\
0.8660      	0.00074\\
1.7320      	0.00367\\
3.4641      	0.02706\\
5.1961      	0.10009\\
6.9282      	0.24436\\
8.6602      	0.45048\\
10.392      	0.70239\\
13.856      	1.57218\\
17.320          2.81098\\
};
\addlegendentry{Proj. onto $\mathcal{V}$}
\addlegendimage{empty legend}
\addlegendentry{\hspace{-.6cm}\textbf{Random Design}}
\addplot [color=red, dashed, line width=1.2pt, mark=square, mark options = solid]
  table[row sep=crcr]{%
0.3464         0.00016\\
0.8660         0.00255\\
1.7320         0.01951\\
3.4641         0.15980\\
5.1961         0.53889\\
6.9282         1.22492\\
8.6602         2.53247\\
10.392         4.20990\\
13.856         9.92737\\
17.320         19.81937\\
};
\addlegendentry{ \emph{Uncon.} beam}
\addplot [color=green, dashed, line width=1.2pt, mark=o, mark options = solid, mark size = 3]
  table[row sep=crcr]{%
0.3464         0.00018\\
0.8660         0.00275\\
1.7320         0.02228\\
3.4641         0.17784\\
5.1961         0.60789\\
6.9282         1.44392\\
8.6602         2.70083\\
10.392         4.89874\\
13.856         11.44478\\
17.320         21.75501\\
};
\addlegendentry{Proj. onto $\mathcal{K}1$}
\addplot [color=blue, dashed, line width=1.2pt, mark=+, mark options = solid, mark size = 3]
  table[row sep=crcr]{%
0.3464         0.00019\\
0.8660      	0.00276\\
1.7320      	0.02251\\
3.4641      	0.17457\\
5.1961      	0.57451\\
6.9282      	1.40897\\
8.6602      	2.98439\\
10.392      	4.72464\\
13.856      	11.49018\\
17.320         22.39948\\
};
\addlegendentry{Proj. onto $\mathcal{K}2$}
\addplot [color=gray, dashed, line width=1.2pt, mark=asterisk, mark options = solid, mark size = 3]
  table[row sep=crcr]{%
0.3464          0.00025\\
0.8660      	0.00443\\
1.7320      	0.03914\\
3.4641      	0.31174\\
5.1961      	1.03910\\
6.9282      	2.53780\\
8.6602      	4.89841\\
10.392      	8.11626\\
13.856      	19.49614\\
17.320          37.75811\\
};
\addlegendentry{Proj. onto $\mathcal{V}$}
\addlegendimage{empty legend}
\addlegendentry{\hspace{-.6cm}\textbf{Directional Design}}
\addplot [color=red, dashdotted, line width=1.2pt, mark=x, mark size=3, mark options={solid}]
  table[row sep=crcr]{%
0.3464                0.0001\\
0.8660      	      0.0008\\
1.7320      	      0.0037\\
3.4641      	      0.0187\\
5.1961      	      0.0523\\
6.9282      	      0.1309\\
8.6602      	      0.3506\\
10.392      	      0.7456\\
13.856      	      2.6302\\
17.320                8.0080\\
};
\addlegendentry{\emph{Uncon.} Beam, $0.5~\mathrm{m}$}
\addplot [color=red, dotted, line width=1.2pt, mark=x, mark size=3, mark options={solid}]
  table[row sep=crcr]{%
0.3464                 0.0002\\
0.8660      	       0.0034\\
1.7320      	       0.0148\\
3.4641      	       0.0463\\
5.1961      	       0.1163\\
6.9282      	       0.2341\\
8.6602      	       0.4214\\
10.392      	       0.5796\\
13.856      	       1.1805\\
17.320                 2.1020\\
};
\addlegendentry{\emph{Uncon.} Beam, $2~\mathrm{m}$}
    
\end{axis}

\begin{axis}[%
width=5.833in,
height=4.375in,
at={(0in,0in)},
scale only axis,
xmin=0,
xmax=1,
ymin=0,
ymax=1,
axis line style={draw=none},
ticks=none,
axis x line*=bottom,
axis y line*=left
]
\end{axis}
\end{tikzpicture}%
}
    \caption[]%
    {PEB as a function of \ac{RIS}-\ac{UE} distance for ideal and designed localization-optimal RIS beams, constrained by the real look-up tables/projection sets of the 4 characterized hardware prototypes.}
    \label{fig:PEBdist}
\end{figure}
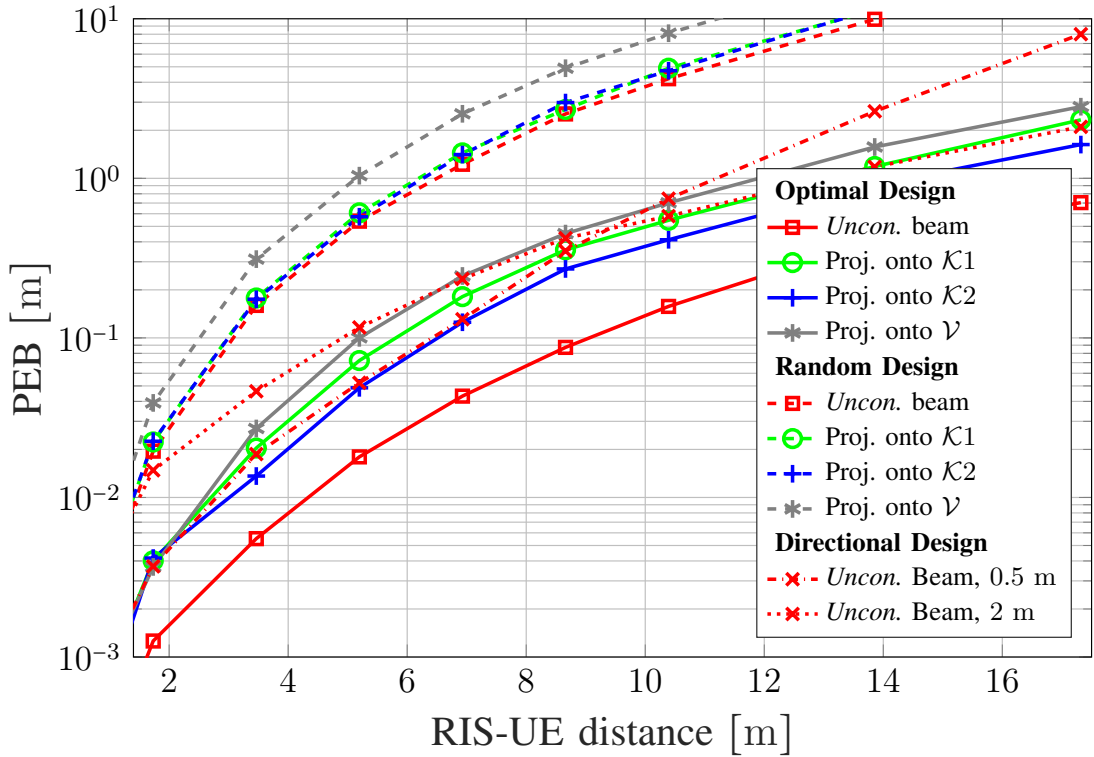

In Fig.~\ref{fig:PEBdist}, we show that \ac{PEB} increases globally as a function of the \ac{RIS}-\ac{UE} distance, as expected. We hence first note that the localization-optimal RIS phase design outperforms the random designs \cite{Rahal_Localization-Optimal_RIS}) for both \emph{constrained} and \emph{unconstrained} beam patterns, whatever the considered range.
Then, among the random phase designs more specifically, the performance of the so-called \emph{unconstrained} beam is only slightly better than that of all the \emph{constrained} beams, which are very close to each other, even if $\mathcal{V}$ looks slightly worse at very first sight (for the same reasons of unfavorable reflection coefficients distribution as before). Overall, this means in practice that the localization performance is not that sensitive to beam fidelity issues caused by beam patterns approximation under real hardware limitations, as long as all the feasible phases/amplitudes of the reflection coefficients are all visited within random \ac{RIS} configuration designs.
Regarding the localization-optimal phase design, which requires a weighed combination of $4$ distinct beams, performance now seems much more sensitive to beam pattern approximation than that in the random design case, while experiencing much higher performance degradation in comparison with the \emph{unconstrained} case. This performance gap looks relatively constant as the \ac{RIS}-\ac{UE} distance increases though. Moreover, as expected, set $\mathcal{K}2$ seems relatively better than $\mathcal{K}1$ which is also better than $\mathcal{V}$. Finally, another set of two curves related to directional \ac{RIS} profile design are shown with different uncertainty sphere radii ($0.5~\mathrm{m}$ and $2~\mathrm{m}$). We Notice that in general, the \ac{PEB} of the directional design lies in between the localization-optimal and the random once. Furthermore, distributing the positions in a smaller uncertainty sphere yields a \ac{PEB} close to that projected onto sets $\mathcal{K}1$ and $\mathcal{K}1$ in the localization-optimal design only at short distances but then degrades quickly as the \ac{RIS}-\ac{UE} distance increases. On the other hand, using the bigger sphere gives the opposite behavior, this is inline with the results presented in~\cite{Rahal_Localization-Optimal_RIS}.

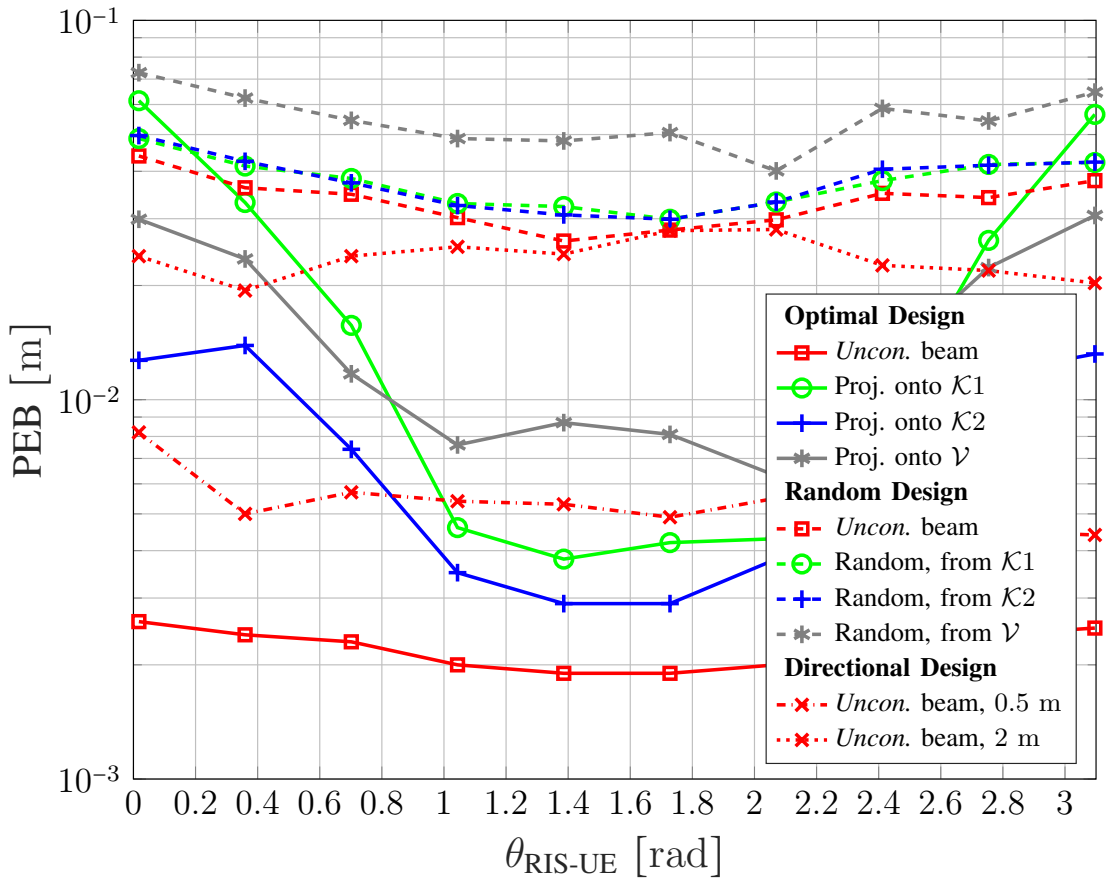
\begin{figure}[ht]
    \centering
    \resizebox{\columnwidth}{!}{
%
\begin{tikzpicture}

\begin{axis}[%
width=4.521in,
height=3.566in,
at={(0.758in,0.481in)},
scale only axis,
xmin=0,
xmax=3.1,
xlabel style={font=\large \color{white!15!black}},
xlabel={$\theta_{\text{RIS-UE}}~[\mathrm{rad}]$},
ymin=0.001,
ymax=0.1,
ymode = log,
ylabel style={font=\large \color{white!15!black}},
ylabel={PEB $[\mathrm{m}]$},
axis background/.style={fill=white},
xmajorgrids,
yminorgrids,
legend style={font = \footnotesize, at={(0.98,0.64)},legend cell align=left, align=left, draw=white!15!black}
]
\addlegendimage{empty legend}
\addlegendentry{\hspace{-.6cm}\textbf{Optimal Design}}
\addplot [color=red, line width=1.2pt, mark=square]
  table[row sep=crcr]{%
0.0175              0.0026\\
0.3595              0.0024\\
0.7016              0.0023\\
1.0437              0.0020\\
1.3858              0.0019\\
1.7279              0.0019\\
2.0700	            0.0020\\
2.4120	            0.0023\\
2.7541	            0.0024\\
3.0962	            0.0025\\
};
\addlegendentry{\emph{Uncon.} beam}

\addplot [color=green, line width=1.2pt, mark=o, mark size = 3]
  table[row sep=crcr]{%
0.0175           0.0614\\
0.3595           0.0331\\
0.7016           0.0157\\
1.0437           0.0046\\
1.3858           0.0038\\
1.7279           0.0042\\
2.0700	         0.0043\\
2.4120	         0.0087\\
2.7541	         0.0263\\
3.0962	         0.0565\\
};
\addlegendentry{Proj. onto $\mathcal{K}1$}

\addplot [color=blue, line width=1.2pt, mark=+, mark size = 3]
  table[row sep=crcr]{%
0.0175          0.0127\\
0.3595          0.0139\\
0.7016          0.0074\\
1.0437          0.0035\\
1.3858          0.0029\\
1.7279          0.0029\\
2.0700	        0.0038\\
2.4120	        0.0066\\
2.7541	        0.0117\\
3.0962	        0.0132\\
};
\addlegendentry{Proj. onto $\mathcal{K}2$}

\addplot [color=gray, line width=1.2pt, mark=asterisk, mark size = 3]
  table[row sep=crcr]{%
0.0175          0.0299\\
0.3595          0.0235\\
0.7016          0.0117\\
1.0437          0.0076\\
1.3858          0.0087\\
1.7279          0.0081\\
2.0700	        0.0063\\
2.4120	        0.0124\\
2.7541	        0.0223\\
3.0962	        0.0306\\
};
\addlegendentry{Proj. onto $\mathcal{V}$}
\addlegendimage{empty legend}
\addlegendentry{\hspace{-.6cm}\textbf{Random Design}}
\addplot [color=red, dashed, line width=1.2pt, mark=square, mark options={solid}]
  table[row sep=crcr]{%
0.0175               0.0439\\
0.3595               0.0362\\
0.7016               0.0348\\
1.0437               0.0302\\
1.3858               0.0262\\
1.7279               0.0280\\
2.0700	             0.0298\\
2.4120	             0.0350\\
2.7541	             0.0341\\
3.0962	             0.0378\\
};
\addlegendentry{\emph{Uncon.} beam}

\addplot [color=green, dashed, line width=1.2pt, mark=o, mark size=3, mark options={solid}]
  table[row sep=crcr]{%
0.0175               0.0488\\
0.3595               0.0413\\
0.7016               0.0383\\
1.0437               0.0329\\
1.3858               0.0323\\
1.7279               0.0299\\
2.0700	             0.0332\\
2.4120	             0.0378\\
2.7541	             0.0417\\
3.0962	             0.0422\\
};
\addlegendentry{Random, from $\mathcal{K}1$}

\addplot [color=blue, dashed, line width=1.2pt, mark=, mark=+, mark size=3, mark options={solid}]
  table[row sep=crcr]{%
0.0175                0.0497\\
0.3595                0.0425\\
0.7016                0.0373\\
1.0437                0.0325\\
1.3858                0.0307\\
1.7279                0.0299\\
2.0700	              0.0332\\
2.4120	              0.0405\\
2.7541	              0.0415\\
3.0962	              0.0423\\
};
\addlegendentry{Random, from $\mathcal{K}2$}

\addplot [color=gray, dashed, line width=1.2pt, mark=asterisk, mark size=3, mark options={solid}]
  table[row sep=crcr]{%
0.0175                0.0728\\
0.3595                0.0624\\
0.7016                0.0545\\
1.0437                0.0488\\
1.3858                0.0481\\
1.7279                0.0506\\
2.0700	              0.0401\\
2.4120	              0.0586\\
2.7541	              0.0543\\
3.0962	              0.0647\\
};
\addlegendentry{Random, from $\mathcal{V}$}
\addlegendimage{empty legend}
\addlegendentry{\hspace{-.6cm}\textbf{Directional Design}}
\addplot [color=red, dashdotted, line width=1.2pt, mark=x, mark size=3, mark options={solid}]
  table[row sep=crcr]{%
0.0175             0.0082\\
0.3595             0.0050\\
0.7016             0.0057\\
1.0437             0.0054\\
1.3858             0.0053\\
1.7279             0.0049\\
2.0700	           0.0055\\
2.4120	           0.0057\\
2.7541	           0.0045\\
3.0962	           0.0044\\
};
\addlegendentry{\emph{Uncon.} beam, $0.5~\mathrm{m}$}
\addplot [color=red, dotted, line width=1.2pt, mark=x, mark size=3, mark options={solid}]
  table[row sep=crcr]{%
0.0175              0.0239\\
0.3595              0.0194\\
0.7016              0.0239\\
1.0437              0.0253\\
1.3858              0.0242\\
1.7279              0.0279\\
2.0700	            0.0281\\
2.4120	            0.0226\\
2.7541	            0.0219\\
3.0962	            0.0203\\
};
\addlegendentry{\emph{Uncon.} beam, $2~\mathrm{m}$}

\end{axis}

\begin{axis}[%
width=5.833in,
height=4.375in,
at={(0in,0in)},
scale only axis,
xmin=0,
xmax=1,
ymin=0,
ymax=1,
axis line style={draw=none},
ticks=none,
axis x line*=bottom,
axis y line*=left
]
\end{axis}
\end{tikzpicture}%
}
    \caption[]%
    {PEB as a function of \ac{RIS}-\ac{UE} azimuth angle for ideal and designed localization-optimal RIS beams, constrained by the real look-up tables/projection sets of the 4 characterized hardware prototypes.}
    \label{fig:PEBaz}
\end{figure}

\begin{figure}[ht]
     \centering
     \resizebox{\columnwidth}{!}{
%
\begin{tikzpicture}

\begin{axis}[%
width=4.521in,
height=3.566in,
at={(0.758in,0.481in)},
scale only axis,
xmin=0,
xmax=3.1,
xlabel style={font=\large \color{white!15!black}},
xlabel={$\phi_{\text{RIS-UE}}~[\mathrm{rad}]$},
ymin=0.001,
ymax=0.1,
ymode = log,
ylabel style={font=\large \color{white!15!black}},
ylabel={PEB $[\mathrm{m}]$},
axis background/.style={fill=white},
xmajorgrids,
yminorgrids,
legend style={font=\footnotesize, at={(0.98,0.64)}, legend cell align=left, align=left, draw=white!15!black}
]
\addlegendimage{empty legend}
\addlegendentry{\hspace{-.6cm}\textbf{Optimal Design}}
\addplot [color=red, line width=1.2pt, mark=square]
  table[row sep=crcr]{%
0.0175      0.00229\\
0.3595      0.00187\\
0.7016      0.00177\\
1.0437      0.00162\\
1.3858      0.00152\\
1.7279      0.00152\\
2.0700      0.00161\\
2.4120      0.00176\\
2.7541      0.00186\\
3.0962      0.00196\\
};
\addlegendentry{\emph{Uncon.} beam}

\addplot [color=green, line width=1.2pt,, mark=o, mark size = 3]
  table[row sep=crcr]{%
0.0175      0.04155\\
0.3595      0.01372\\
0.7016      0.00399\\
1.0437      0.00391\\
1.3858      0.00344\\
1.7279      0.00351\\
2.0700      0.00449\\
2.4120      0.00684\\
2.7541      0.00436\\
3.0962      0.06917\\
};
\addlegendentry{Proj. onto $\mathcal{K}1$}

\addplot [color=blue, line width=1.2pt, mark=+, mark size = 3]
  table[row sep=crcr]{%
0.0175          0.07857\\
0.3595          0.00992\\
0.7016          0.00381\\
1.0437          0.00258\\
1.3858          0.00251\\
1.7279          0.00258\\
2.0700          0.00321\\
2.4120          0.00408\\
2.7541          0.00655\\
3.0962          0.01702\\
};
\addlegendentry{Proj. onto $\mathcal{K}2$}

\addplot [color=gray, line width=1.2pt, mark=asterisk, mark size = 3]
  table[row sep=crcr]{%
0.0175          0.24236\\
0.3595          0.00910\\
0.7016          0.00525\\
1.0437          0.00437\\
1.3858          0.00444\\
1.7279          0.00456\\
2.0700          0.00428\\
2.4120          0.00489\\
2.7541          0.01210\\
3.0962          0.21846\\
};
\addlegendentry{Proj. onto $\mathcal{V}$}
\addlegendimage{empty legend}
\addlegendentry{\hspace{-.6cm}\textbf{Random Design}}
\addplot [color=red, dashed, line width=1.2pt, mark=square, mark options = solid]
  table[row sep=crcr]{%
0.0175         0.04253\\
0.3595      0.03627\\
0.7016      0.03392\\
1.0437      0.03048\\
1.3858      0.02643\\
1.7279         0.02608\\
2.0700     	0.02865\\
2.4120     	0.03387\\
2.7541     	0.03642\\
3.0962     	0.03808\\
};
\addlegendentry{\emph{Uncon.} beam}

\addplot [color=green, dashed, line width=1.2pt, mark=o, mark options = solid, mark size = 3]
  table[row sep=crcr]{%
0.0175         0.04648\\
0.3595      0.03899\\
0.7016      0.03739\\
1.0437      0.03329\\
1.3858      0.03018\\
1.7279         0.02924\\
2.0700     	0.03222\\
2.4120     	0.03652\\
2.7541     	0.04079\\
3.0962     	0.04342\\
};
\addlegendentry{Random, from $\mathcal{K}1$}

\addplot [color=blue, dashed, line width=1.2pt, mark=+, mark options = solid, mark size = 3]
  table[row sep=crcr]{%
0.0175      0.04756\\
0.3595   0.03937\\
0.7016   0.03629\\
1.0437   0.03285\\
1.3858   0.02885\\
1.7279      0.02976\\
2.0700   0.03207\\
2.4120   0.03771\\
2.7541   0.04162\\
3.0962   0.04375\\
};
\addlegendentry{Random, from $\mathcal{K}2$}

\addplot [color=gray, dashed, line width=1.2pt, mark=asterisk, mark options = solid, mark size = 3]
  table[row sep=crcr]{%
0.0175         0.08832\\
0.3595      0.06956\\
0.7016      0.06599\\
1.0437      0.05770\\
1.3858      0.05184\\
1.7279         0.04844\\
2.0700      0.05449\\
2.4120      0.06509\\
2.7541      0.07371\\
3.0962      0.07413\\
};
\addlegendentry{Random, from $\mathcal{V}$}
\addlegendimage{empty legend}
\addlegendentry{\hspace{-.6cm}\textbf{Directional Design}}
\addplot [color=red, dashdotted, line width=1.2pt, mark=x, mark size=3, mark options={solid}]
  table[row sep=crcr]{%
0.0175            0.0084\\
0.3595            0.0050\\
0.7016            0.0059\\
1.0437            0.0054\\
1.3858            0.0051\\
1.7279            0.0047\\
2.0700   	      0.0053\\
2.4120   	      0.0053\\
2.7541   	      0.0046\\
3.0962   	      0.0041\\
};
\addlegendentry{\emph{Uncon.} beam}
\addplot [color=red, dotted, line width=1.2pt, mark=x, mark size=3, mark options={solid}]
  table[row sep=crcr]{%
0.0175                  0.0645\\
0.3595                  0.0270\\
0.7016                  0.0237\\
1.0437                  0.0275\\
1.3858                  0.0297\\
1.7279                  0.0221\\
2.0700   	            0.0259\\
2.4120   	            0.0202\\
2.7541   	            0.0158\\
3.0962   	            0.0158\\
};
\addlegendentry{\emph{Uncon.} beam}

\end{axis}
\begin{axis}[%
width=5.833in,
height=4.375in,
at={(0in,0in)},
scale only axis,
xmin=0,
xmax=1,
ymin=0,
ymax=1,
axis line style={draw=none},
ticks=none,
axis x line*=bottom,
axis y line*=left
]
\end{axis}
\end{tikzpicture}%
}
    \caption[]%
    {PEB as a function of \ac{RIS}-\ac{UE} elevation angle for ideal and designed localization-optimal RIS beams, constrained by the real look-up tables/projection sets of the 4 characterized hardware prototypes.}
    \label{fig:PEBel}
\end{figure}
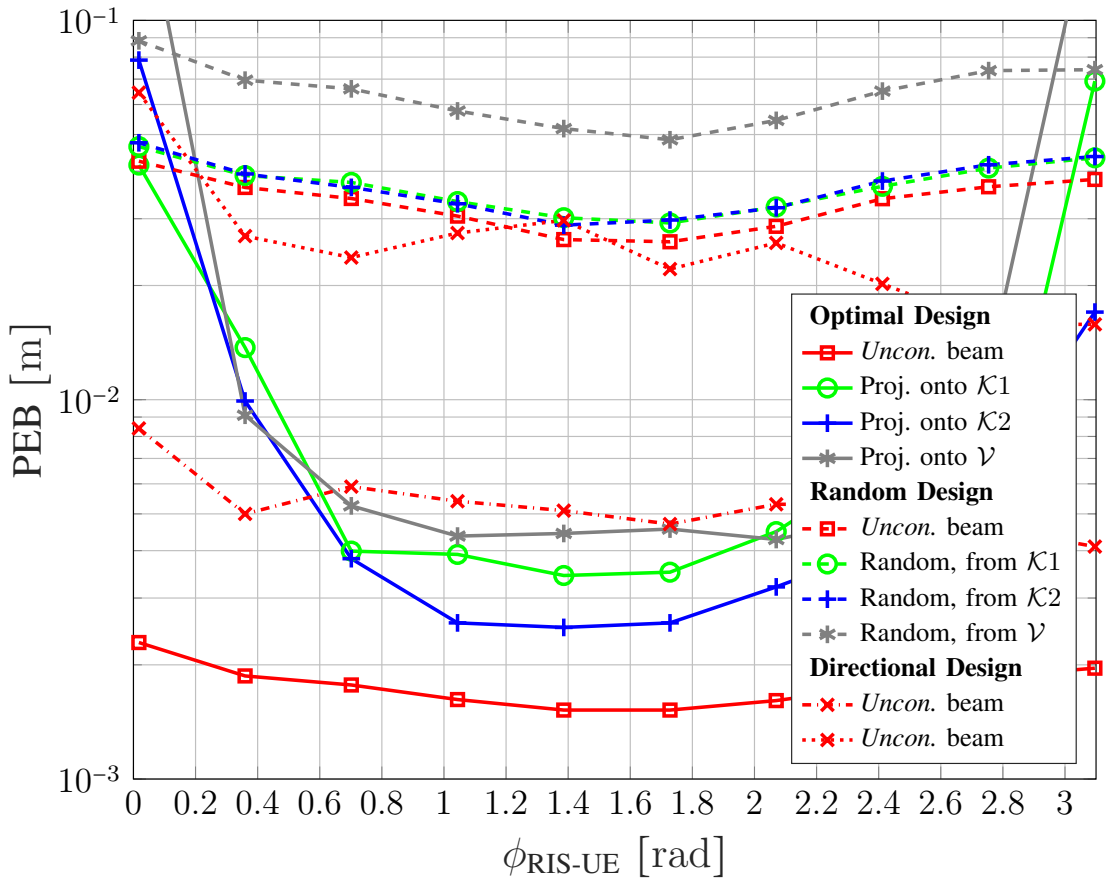

Fig.~\ref{fig:PEBaz} shows the PEB evolution as a function of the azimuth angle (i.e., the \ac{RIS}-\ac{UE} distance and elevation being fixed), where one can observe the same trends and ranking as before with the \ac{RIS}-\ac{UE} range. 
Whatever the setting, the \ac{PEB} is also mainly better in the inner part of the spanned angular interval, illustrating typical geometric effects as we get away from the boresight, regardless of beam approximation. Since the \ac{RIS}-\ac{UE} distance is short, utilizing a directional phase design with a smaller sphere yields a better performance than a bigger one which in turn almost performs as the random design.
Likewise, Fig.~\ref{fig:PEBel} shows similar \ac{PEB} curves as a function of the elevation angle (i.e., the \ac{RIS}-\ac{UE} distance and azimuth being fixed), where performance is still better in the inner part of the spanned angular intervals, illustrating typical effects as we get away from the boresight. The gap between the \ac{PEB} performance achieved with an \emph{unconstrained} beam pattern and its \emph{constrained} approximated variants seems all the more critical in those outer angular zones.
%
\section*{Conclusion and Future Work}
%
In this study, we presented a low-complexity technique for optimizing the complex configuration of \acp{RIS} so as to create various beam patterns, while considering real hardware limitations. The proposed method utilizes a pre-determined lookup table of possible \ac{RIS} element reflection coefficients, characterized out of real measurements. 
Applying the proposed beam generation strategy into the concrete context of \ac{NF} \ac{NLoS} localization, the performance degradation induced by beams approximation has then been evaluated in terms of both beam fidelity and \ac{PEB}, and further benchmarked for different a priori \ac{RIS} control strategies (incl. random, directional and localization-optimal designs). First numerical simulations regarding beam fidelity emphasize the dominating influence of both phase quantization levels (and the span of practically valid phases) and power losses regarding the element-wise complex reflection coefficients on the power gain of the \ac{RIS} beam peak in the desired \ac{UE} direction, as well as on the existence of unwanted secondary lobes. Other simulations performed in a canonical scenario also reveal the effects of beam synthesis and their relative power losses on the localization performance of different \ac{RIS} prototypes. The \ac{PEB} metric was utilized for that manner and the positioning was evaluated versus both \ac{RIS}-\ac{UE} distance and angles (i.e., both azimuth and elevation) to ensure that the study covers all the possible dimensions. As expected, constraining the beam synthesis to \ac{RIS} designs with certain limits degrade the positioning performance across all dimensions with various \ac{RIS} profile designs.

Future works should investigate more \ac{RIS} prototypes by testing their different design characteristics based on localization performance. Extending the results to state-of-the-art as well as novel localization algorithms is also a possible path where the gaps in the performance of the \ac{RIS} devices can be analyzed and fixed.

\

%
%

%
%
%



\section*{Acknowledgements}
The authors would like to warmly thank Dr. Antonio Clemente, from CEA-Leti, France, Dr. Philippe Ratajczak, from Orange Innovation Networks, France, for sharing the look-up tables associated to their R-RIS hardware prototypes in the context of RISE-6G project, as well as for the fruitful technical discussions and the relevant advice regarding this paper.

\section*{Funding}
This work has been supported, in part, by the EU H2020
RISE-6G project under grant 101017011 and by the MSCA-IF
grant 888913 (OTFS-RADCOM).

\section*{Abbreviations}
\begin{itemize}
\item BS: Base Station
\item DoD: Direction of Departure
\item FIM: Fisher Information Matrix
\item LoS: Line-of-Sight
\item LMI: Linear Matrix Inequalities
\item MC: Multi-Carrier
\item MIMO: Multiple Inputs Multiple Outputs
\item NF: Nearfield
\item NLoS: Non-Line-of-Sight
\item OEB: Orientation Error Bound
\item PEB: Position Error Bound
\item PSD: Positive Semi Definite
\item RIS: Reconfigurable Intelligent Surface
\item QoS: Quality of Service
\item RX: Receiver
\item SDP: Semi Definite Program
\item S(I)NR: Signal-to-(Interference-and-)Noise-Ratio
\item SISO: Single-Input-Single-Output
\item SRE: Smart Radio Environment
\item UE: User Equipment
\item TDoA: Time Difference of Arrival
\item TX: Transmitter
\end{itemize}


\bibliographystyle{bmc-mathphys} 
\bibliography{bmc_article}      


\begin{thebibliography}{47}
\ifx \bisbn   \undefined \def \bisbn  #1{ISBN #1}\fi
\ifx \binits  \undefined \def \binits#1{#1}\fi
\ifx \bauthor  \undefined \def \bauthor#1{#1}\fi
\ifx \batitle  \undefined \def \batitle#1{#1}\fi
\ifx \bjtitle  \undefined \def \bjtitle#1{#1}\fi
\ifx \bvolume  \undefined \def \bvolume#1{\textbf{#1}}\fi
\ifx \byear  \undefined \def \byear#1{#1}\fi
\ifx \bissue  \undefined \def \bissue#1{#1}\fi
\ifx \bfpage  \undefined \def \bfpage#1{#1}\fi
\ifx \blpage  \undefined \def \blpage #1{#1}\fi
\ifx \burl  \undefined \def \burl#1{\textsf{#1}}\fi
\ifx \doiurl  \undefined \def \doiurl#1{\textsf{#1}}\fi
\ifx \betal  \undefined \def \betal{\textit{et al.}}\fi
\ifx \binstitute  \undefined \def \binstitute#1{#1}\fi
\ifx \binstitutionaled  \undefined \def \binstitutionaled#1{#1}\fi
\ifx \bctitle  \undefined \def \bctitle#1{#1}\fi
\ifx \beditor  \undefined \def \beditor#1{#1}\fi
\ifx \bpublisher  \undefined \def \bpublisher#1{#1}\fi
\ifx \bbtitle  \undefined \def \bbtitle#1{#1}\fi
\ifx \bedition  \undefined \def \bedition#1{#1}\fi
\ifx \bseriesno  \undefined \def \bseriesno#1{#1}\fi
\ifx \blocation  \undefined \def \blocation#1{#1}\fi
\ifx \bsertitle  \undefined \def \bsertitle#1{#1}\fi
\ifx \bsnm \undefined \def \bsnm#1{#1}\fi
\ifx \bsuffix \undefined \def \bsuffix#1{#1}\fi
\ifx \bparticle \undefined \def \bparticle#1{#1}\fi
\ifx \barticle \undefined \def \barticle#1{#1}\fi
\ifx \bconfdate \undefined \def \bconfdate #1{#1}\fi
\ifx \botherref \undefined \def \botherref #1{#1}\fi
\ifx \url \undefined \def \url#1{\textsf{#1}}\fi
\ifx \bchapter \undefined \def \bchapter#1{#1}\fi
\ifx \bbook \undefined \def \bbook#1{#1}\fi
\ifx \bcomment \undefined \def \bcomment#1{#1}\fi
\ifx \oauthor \undefined \def \oauthor#1{#1}\fi
\ifx \citeauthoryear \undefined \def \citeauthoryear#1{#1}\fi
\ifx \endbibitem  \undefined \def \endbibitem {}\fi
\ifx \bconflocation  \undefined \def \bconflocation#1{#1}\fi
\ifx \arxivurl  \undefined \def \arxivurl#1{\textsf{#1}}\fi
\csname PreBibitemsHook\endcsname

\bibitem{huang2019reconfigurable}
\begin{barticle}
\bauthor{\bsnm{Huang}, \binits{C.}},
\bauthor{\bsnm{Zappone}, \binits{A.}},
\bauthor{\bsnm{Alexandropoulos}, \binits{G.C.}},
\bauthor{\bsnm{Debbah}, \binits{M.}},
\bauthor{\bsnm{Yuen}, \binits{C.}}:
\batitle{Reconfigurable intelligent surfaces for energy efficiency in wireless
  communication}.
\bjtitle{IEEE Trans. Wireless Commun.}
\bvolume{18}(\bissue{8}),
\bfpage{4157}--\blpage{4170}
(\byear{2019})
\end{barticle}
\endbibitem

\bibitem{qignqingwu2019}
\begin{barticle}
\bauthor{\bsnm{Wu}, \binits{Q.}},
\bauthor{\bsnm{Zhang}, \binits{R.}}:
\batitle{Intelligent reflecting surface enhanced wireless network via joint
  active and passive beamforming}.
\bjtitle{IEEE Trans. Wireless Commun.}
\bvolume{18}(\bissue{11}),
\bfpage{5394}--\blpage{5409}
(\byear{2019})
\end{barticle}
\endbibitem

\bibitem{RISE6G_COMMAG}
\begin{barticle}
\bauthor{\bsnm{Calvanese~Strinati}, \binits{E.}},
\bauthor{\bsnm{Alexandropoulos}, \binits{G.C.}},
\bauthor{\bsnm{Wymeersch}, \binits{H.}},
\bauthor{\bsnm{Denis}, \binits{B.}},
\bauthor{\bsnm{Sciancalepore}, \binits{V.}},
\bauthor{\bsnm{D'Errico}, \binits{R.}},
\bauthor{\bsnm{Clemente}, \binits{A.}},
\bauthor{\bsnm{Phan-Huy}, \binits{D.-T.}},
\bauthor{\bsnm{Carvalho}, \binits{E.D.}},
\bauthor{\bsnm{Popovski}, \binits{P.}}:
\batitle{Reconfigurable, intelligent, and sustainable wireless environments for
  {6G} smart connectivity}.
\bjtitle{IEEE Commun. Mag.}
\bvolume{59}(\bissue{10}),
\bfpage{99}--\blpage{105}
(\byear{2021})
\end{barticle}
\endbibitem

\bibitem{Kamran_Leveraging}
\begin{botherref}
\oauthor{\bsnm{Keykhosravi}, \binits{K.}},
\oauthor{\bsnm{Denis}, \binits{B.}},
\oauthor{\bsnm{Alexandropoulos}, \binits{G.C.}},
\oauthor{\bsnm{He}, \binits{Z.S.}},
\oauthor{\bsnm{Albanese}, \binits{A.}},
\oauthor{\bsnm{Sciancalepore}, \binits{V.}},
\oauthor{\bsnm{Wymeersch}, \binits{H.}}:
Leveraging RIS-Enabled Smart Signal Propagation for Solving Infeasible
  Localization Problems
(2022).
doi:\doiurl{10.48550/arXiv.2204.11538}
\end{botherref}
\endbibitem

\bibitem{Molisch_HBF_2017_all}
\begin{barticle}
\bauthor{\bsnm{Molisch}, \binits{A.F.}},
\bauthor{\bsnm{Ratnam}, \binits{V.V.}},
\bauthor{\bsnm{Han}, \binits{S.}},
\bauthor{\bsnm{Li}, \binits{Z.}},
\bauthor{\bsnm{Nguyen}, \binits{S.L.H.}},
\bauthor{\bsnm{Li}, \binits{L.}},
\bauthor{\bsnm{Haneda}, \binits{K.}}:
\batitle{Hybrid beamforming for massive {MIMO}: {A} survey}.
\bjtitle{{IEEE} {Commun.} {M}ag.}
\bvolume{55}(\bissue{9}),
\bfpage{134}--\blpage{141}
(\byear{2017})
\end{barticle}
\endbibitem

\bibitem{alexandg_2021}
\begin{barticle}
\bauthor{\bsnm{Alexandropoulos}, \binits{G.C.}},
\bauthor{\bsnm{Shlezinger}, \binits{N.}},
\bauthor{\bparticle{del} \bsnm{Hougne}, \binits{P.}}:
\batitle{Reconfigurable intelligent surfaces for rich scattering wireless
  communications: {R}ecent experiments, challenges, and opportunities}.
\bjtitle{IEEE Commun. Mag.}
\bvolume{59}(\bissue{6}),
\bfpage{28}--\blpage{34}
(\byear{2021})
\end{barticle}
\endbibitem

\bibitem{Abeywickrama_2020}
\begin{barticle}
\bauthor{\bsnm{Abeywickrama}, \binits{S.}},
\bauthor{\bsnm{Zhang}, \binits{R.}},
\bauthor{\bsnm{Wu}, \binits{Q.}},
\bauthor{\bsnm{Yue}, \binits{C.}}:
\batitle{Intelligent reflecting surface: {P}ractical phase shift model and
  beamforming optimization}.
\bjtitle{IEEE Trans. Commun.}
\bvolume{68}(\bissue{9}),
\bfpage{5849}--\blpage{5863}
(\byear{2020})
\end{barticle}
\endbibitem

\bibitem{Larsson_2021}
\begin{barticle}
\bauthor{\bsnm{Long}, \binits{R.}},
\bauthor{\bsnm{Liang}, \binits{Y.-C.}},
\bauthor{\bsnm{Pei}, \binits{Y.}},
\bauthor{\bsnm{Larsson}, \binits{E.G.}}:
\batitle{Active reconfigurable intelligent surface-aided wireless
  communications}.
\bjtitle{IEEE Trans. Wireless Commun.}
\bvolume{20}(\bissue{8}),
\bfpage{4962}--\blpage{4975}
(\byear{2021})
\end{barticle}
\endbibitem

\bibitem{RIS_Impairments}
\begin{bchapter}
\bauthor{\bsnm{Hu}, \binits{S.}},
\bauthor{\bsnm{Rusek}, \binits{F.}},
\bauthor{\bsnm{Edfors}, \binits{O.}}:
\bctitle{Capacity degradation with modeling hardware impairment in large
  intelligent surface}.
In: \bbtitle{Proc. IEEE Global Communications Conference (GLOBECOM)}
(\byear{2018})
\end{bchapter}
\endbibitem

\bibitem{shen2020beamforming}
\begin{barticle}
\bauthor{\bsnm{Shen}, \binits{H.}},
\bauthor{\bsnm{Xu}, \binits{W.}},
\bauthor{\bsnm{Gong}, \binits{S.}},
\bauthor{\bsnm{Zhao}, \binits{C.}},
\bauthor{\bsnm{Ng}, \binits{D.W.K.}}:
\batitle{Beamforming optimization for {IRS}-aided communications with
  transceiver hardware impairments}.
\bjtitle{IEEE Trans. Commun.}
\bvolume{69}(\bissue{2}),
\bfpage{1214}--\blpage{1227}
(\byear{2020})
\end{barticle}
\endbibitem

\bibitem{Rahal_EuCNC22}
\begin{bchapter}
\bauthor{\bsnm{Rahal}, \binits{M.}},
\bauthor{\bsnm{Denis}, \binits{B.}},
\bauthor{\bsnm{Keykhosravi}, \binits{K.}},
\bauthor{\bsnm{Keskin}, \binits{M.F.}},
\bauthor{\bsnm{Uguen}, \binits{B.}},
\bauthor{\bsnm{Alexandropoulos}, \binits{G.C.}},
\bauthor{\bsnm{Wymeersch}, \binits{H.}}:
\bctitle{Arbitrary beam pattern approximation via riss with measured element
  responses}.
In: \bbtitle{Joint European Conference on Networks and Communications \& 6G
  Summit (EuCNC/6G Summit)},
pp. \bfpage{506}--\blpage{511}
(\byear{2022}).
doi:\doiurl{10.1109/EuCNC/6GSummit54941.2022.9815624}
\end{bchapter}
\endbibitem

\bibitem{Elzanaty2021}
\begin{barticle}
\bauthor{\bsnm{Elzanaty}, \binits{A.}},
\bauthor{\bsnm{Guerra}, \binits{A.}},
\bauthor{\bsnm{Guidi}, \binits{F.}},
\bauthor{\bsnm{Alouini}, \binits{M.-S.}}:
\batitle{Reconfigurable intelligent surfaces for localization: Position and
  orientation error bounds}.
\bjtitle{IEEE Transactions on Signal Processing}
\bvolume{69},
\bfpage{5386}--\blpage{5402}
(\byear{2021}).
doi:\doiurl{10.1109/TSP.2021.3101644}
\end{barticle}
\endbibitem

\bibitem{Rahal_Localization-Optimal_RIS}
\begin{bchapter}
\bauthor{\bsnm{Rahal}, \binits{M.}},
\bauthor{\bsnm{Denis}, \binits{B.}},
\bauthor{\bsnm{Keykhosravi}, \binits{K.}},
\bauthor{\bsnm{Keskin}, \binits{M.F.}},
\bauthor{\bsnm{Uguen}, \binits{B.}},
\bauthor{\bsnm{Wymeersch}, \binits{H.}}:
\bctitle{Constrained ris phase profile optimization and time sharing for
  near-field localization}.
In: \bbtitle{IEEE 95th Vehicular Technology Conference (VTC-Spring)},
pp. \bfpage{1}--\blpage{6}
(\byear{2022}).
doi:\doiurl{10.1109/VTC2022-Spring54318.2022.9860413}
\end{bchapter}
\endbibitem

\bibitem{keykhosravi2021siso}
\begin{bchapter}
\bauthor{\bsnm{Keykhosravi}, \binits{K.}},
\bauthor{\bsnm{Keskin}, \binits{M.F.}},
\bauthor{\bsnm{Seco-Granados}, \binits{G.}},
\bauthor{\bsnm{Wymeersch}, \binits{H.}}:
\bctitle{{SISO} {RIS}-enabled joint {3D} downlink localization and
  synchronization}.
In: \bbtitle{Proc. IEEE International Conference on Communications (ICC)}
(\byear{2021})
\end{bchapter}
\endbibitem

\bibitem{rahal2021ris}
\begin{bchapter}
\bauthor{\bsnm{Rahal}, \binits{M.}},
\bauthor{\bsnm{Denis}, \binits{B.}},
\bauthor{\bsnm{Keykhosravi}, \binits{K.}},
\bauthor{\bsnm{Uguen}, \binits{B.}},
\bauthor{\bsnm{Wymeersch}, \binits{H.}}:
\bctitle{{RIS}-enabled localization continuity under near-field conditions}.
In: \bbtitle{Proc. IEEE 22nd International Workshop on Signal Processing
  Advances in Wireless Communications (SPAWC)}
(\byear{2021}).
doi:\doiurl{10.1109/SPAWC51858.2021.9593200}
\end{bchapter}
\endbibitem

\bibitem{fara2021prototype}
\begin{botherref}
\oauthor{\bsnm{Fara}, \binits{R.}},
\oauthor{\bsnm{Ratajczak}, \binits{P.}},
\oauthor{\bsnm{Huy}, \binits{D.-T.P.}},
\oauthor{\bsnm{Ourir}, \binits{A.}},
\oauthor{\bsnm{Di~Renzo}, \binits{M.}},
\oauthor{\bsnm{De~Rosny}, \binits{J.}}:
A prototype of reconfigurable intelligent surface with continuous control of
  the reflection phase
(2021).
doi:\doiurl{10.48550/arXiv.2105.11862}
\end{botherref}
\endbibitem

\bibitem{DiPalma_2017}
\begin{barticle}
\bauthor{\bsnm{Di~Palma}, \binits{L.}},
\bauthor{\bsnm{Clemente}, \binits{A.}},
\bauthor{\bsnm{Dussopt}, \binits{L.}},
\bauthor{\bsnm{Sauleau}, \binits{R.}},
\bauthor{\bsnm{Potier}, \binits{P.}},
\bauthor{\bsnm{Pouliguen}, \binits{P.}}:
\batitle{Circularly-polarized reconfigurable transmitarray in {K}a-band with
  beam scanning and polarization switching capabilities}.
\bjtitle{IEEE Trans. Antennas Prop.}
\bvolume{65}(\bissue{2}),
\bfpage{529}--\blpage{540}
(\byear{2017}).
doi:\doiurl{10.1109/TAP.2016.2633067}
\end{barticle}
\endbibitem

\bibitem{GNW_Prototype}
\begin{barticle}
\bauthor{\bsnm{Gros}, \binits{J.-B.}},
\bauthor{\bsnm{Popov}, \binits{V.}},
\bauthor{\bsnm{Odit}, \binits{M.A.}},
\bauthor{\bsnm{Lenets}, \binits{V.}},
\bauthor{\bsnm{Lerosey}, \binits{G.}}:
\batitle{A reconfigurable intelligent surface at mmwave based on a binary phase
  tunable metasurface}.
\bjtitle{IEEE Open Journal of the Communications Society}
\bvolume{2},
\bfpage{1055}--\blpage{1064}
(\byear{2021}).
doi:\doiurl{10.1109/OJCOMS.2021.3076271}
\end{barticle}
\endbibitem

\bibitem{monopulse_review}
\begin{barticle}
\bauthor{\bsnm{Nickel}, \binits{U.}}:
\batitle{Overview of generalized monopulse estimation}.
\bjtitle{IEEE Trans. Aerosp. Electron. Syst.}
\bvolume{21}(\bissue{6}),
\bfpage{27}--\blpage{56}
(\byear{2006})
\end{barticle}
\endbibitem

\bibitem{phasedArray_99}
\begin{barticle}
\bauthor{\bsnm{Agrawal}, \binits{A.K.}},
\bauthor{\bsnm{Holzman}, \binits{E.L.}}:
\batitle{Beamformer architectures for active phased-array radar antennas}.
\bjtitle{IEEE Trans. Antennas Prop.}
\bvolume{47}(\bissue{3}),
\bfpage{432}--\blpage{442}
(\byear{1999}).
doi:\doiurl{10.1109/8.768777}
\end{barticle}
\endbibitem

\bibitem{phasedArray_2016}
\begin{barticle}
\bauthor{\bsnm{Talisa}, \binits{S.H.}},
\bauthor{\bsnm{O'Haver}, \binits{K.W.}},
\bauthor{\bsnm{Comberiate}, \binits{T.M.}},
\bauthor{\bsnm{Sharp}, \binits{M.D.}},
\bauthor{\bsnm{Somerlock}, \binits{O.F.}}:
\batitle{Benefits of digital phased array radars}.
\bjtitle{Proc. IEEE}
\bvolume{104}(\bissue{3}),
\bfpage{530}--\blpage{543}
(\byear{2016}).
doi:\doiurl{10.1109/JPROC.2016.2515842}
\end{barticle}
\endbibitem

\bibitem{abrardo2020intelligent}
\begin{botherref}
\oauthor{\bsnm{Abrardo}, \binits{A.}},
\oauthor{\bsnm{Dardari}, \binits{D.}},
\oauthor{\bsnm{Di~Renzo}, \binits{M.}}:
Intelligent reflecting surfaces: Sum-rate optimization based on statistical
  position information.
IEEE Transactions on Communications, July
(2021)
\end{botherref}
\endbibitem

\bibitem{hu2020location}
\begin{barticle}
\bauthor{\bsnm{Hu}, \binits{X.}},
\bauthor{\bsnm{Zhong}, \binits{C.}},
\bauthor{\bsnm{Zhang}, \binits{Y.}},
\bauthor{\bsnm{Chen}, \binits{X.}},
\bauthor{\bsnm{Zhang}, \binits{Z.}}:
\batitle{Location information aided multiple intelligent reflecting surface
  systems}.
\bjtitle{IEEE Transactions on Communications}
\bvolume{68}(\bissue{12}),
\bfpage{7948}--\blpage{7962}
(\byear{2020})
\end{barticle}
\endbibitem

\bibitem{wymeersch_beyond_2020}
\begin{bchapter}
\bauthor{\bsnm{Wymeersch}, \binits{H.}},
\bauthor{\bsnm{Denis}, \binits{B.}}:
\bctitle{Beyond {5G} {Wireless} {Localization} with {Reconfigurable}
  {Intelligent} {Surfaces}}.
In: \bbtitle{{IEEE} {International} {Conference} on {Communications} ({ICC})},
pp. \bfpage{1}--\blpage{6}
(\byear{2020}).
doi:\doiurl{10.1109/ICC40277.2020.9148744}
\end{bchapter}
\endbibitem

\bibitem{Rinchi22}
\begin{barticle}
\bauthor{\bsnm{Rinchi}, \binits{O.}},
\bauthor{\bsnm{Elzanaty}, \binits{A.}},
\bauthor{\bsnm{Alouini}, \binits{M.-S.}}:
\batitle{Compressive near-field localization for multipath ris-aided
  environments}.
\bjtitle{IEEE Communications Letters}
\bvolume{26}(\bissue{6}),
\bfpage{1268}--\blpage{1272}
(\byear{2022}).
doi:\doiurl{10.1109/LCOMM.2022.3151036}
\end{barticle}
\endbibitem

\bibitem{Palmucci22}
\begin{botherref}
\oauthor{\bsnm{Palmucci}, \binits{S.}},
\oauthor{\bsnm{Guerra}, \binits{A.}},
\oauthor{\bsnm{Abrardo}, \binits{A.}},
\oauthor{\bsnm{Dardari}, \binits{D.}}:
RIS-aided User Tracking in Near-Field MIMO Systems: Joint Precoding Design and
  RIS Optimization
(2022).
doi:\doiurl{10.48550/arxiv.2212.07333}
\end{botherref}
\endbibitem

\bibitem{Pan22}
\begin{botherref}
\oauthor{\bsnm{Pan}, \binits{Y.}},
\oauthor{\bsnm{Pan}, \binits{C.}},
\oauthor{\bsnm{Jin}, \binits{S.}},
\oauthor{\bsnm{Wang}, \binits{J.}}:
RIS-Aided Near-Field Localization and Channel Estimation for the Sub-Terahertz
  System.
arXiv
(2022).
doi:\doiurl{10.48550/ARXIV.2208.11343}
\end{botherref}
\endbibitem

\bibitem{Ghazalian22}
\begin{bchapter}
\bauthor{\bsnm{Ghazalian}, \binits{R.}},
\bauthor{\bsnm{Keykhosravi}, \binits{K.}},
\bauthor{\bsnm{Chen}, \binits{H.}},
\bauthor{\bsnm{Wymeersch}, \binits{H.}},
\bauthor{\bsnm{Jäntti}, \binits{R.}}:
\bctitle{Bi-static sensing for near-field ris localization}.
In: \bbtitle{IEEE Global Communications Conference (GLOBECOM)},
pp. \bfpage{6457}--\blpage{6462}
(\byear{2022}).
doi:\doiurl{10.1109/GLOBECOM48099.2022.10001209}
\end{bchapter}
\endbibitem

\bibitem{Mei22}
\begin{barticle}
\bauthor{\bsnm{Mei}, \binits{P.}},
\bauthor{\bsnm{Cai}, \binits{Y.}},
\bauthor{\bsnm{Zhao}, \binits{K.}},
\bauthor{\bsnm{Ying}, \binits{Z.}},
\bauthor{\bsnm{Pedersen}, \binits{G.F.}},
\bauthor{\bsnm{Lin}, \binits{X.Q.}},
\bauthor{\bsnm{Zhang}, \binits{S.}}:
\batitle{On the study of reconfigurable intelligent surfaces in the near-field
  region}.
\bjtitle{IEEE Transactions on Antennas and Propagation}
\bvolume{70}(\bissue{10}),
\bfpage{8718}--\blpage{8728}
(\byear{2022}).
doi:\doiurl{10.1109/TAP.2022.3147533}
\end{barticle}
\endbibitem

\bibitem{alexandropoulos_near-field_2022}
\begin{bchapter}
\bauthor{\bsnm{Alexandropoulos}, \binits{G.C.}},
\bauthor{\bsnm{Jamali}, \binits{V.}},
\bauthor{\bsnm{Schober}, \binits{R.}},
\bauthor{\bsnm{Poor}, \binits{H.V.}}:
\bctitle{Near-field hierarchical beam management for {RIS}-enabled millimeter
  wave multi-antenna systems}.
In: \bbtitle{Proc. IEEE SAM},
pp. \bfpage{460}--\blpage{464}
(\byear{2022})
\end{bchapter}
\endbibitem

\bibitem{liu_low-overhead_2022}
\begin{botherref}
\oauthor{\bsnm{Liu}, \binits{W.}},
\oauthor{\bsnm{Pan}, \binits{C.}},
\oauthor{\bsnm{Ren}, \binits{H.}},
\oauthor{\bsnm{Shu}, \binits{F.}},
\oauthor{\bsnm{Jin}, \binits{S.}},
\oauthor{\bsnm{Wang}, \binits{J.}}:
Low-overhead Beam Training Scheme for Extremely Large-Scale {RIS} in
  Near-field.
arXiv:2211.15910
(2022)
\end{botherref}
\endbibitem

\bibitem{chung_location-aware_2021}
\begin{botherref}
\oauthor{\bsnm{Chung}, \binits{H.}},
\oauthor{\bsnm{Kim}, \binits{S.}}:
Location-aware Channel Estimation for {RIS}-aided {mmWave} {MIMO} Systems via
  Atomic Norm Minimization.
arXiv:2107.09222
(2021)
\end{botherref}
\endbibitem

\bibitem{chen-hu_differential_2022}
\begin{barticle}
\bauthor{\bsnm{Chen-Hu}, \binits{K.}},
\bauthor{\bsnm{Alexandropoulos}, \binits{G.C.}},
\bauthor{\bsnm{Armada}, \binits{A.g.}}:
\batitle{Differential data-aided beam training for {RIS}-empowered
  multi-antenna communications}.
\bjtitle{IEEE Access}
\bvolume{10},
\bfpage{113200}--\blpage{113213}
(\byear{2022})
\end{barticle}
\endbibitem

\bibitem{palmucci_ris-aided_2022}
\begin{botherref}
\oauthor{\bsnm{Palmucci}, \binits{S.}},
\oauthor{\bsnm{Guerra}, \binits{A.}},
\oauthor{\bsnm{Abrardo}, \binits{A.}},
\oauthor{\bsnm{Dardari}, \binits{D.}}:
{RIS}-aided User Tracking in Near-Field {MIMO} Systems: {Joint} Precoding
  Design and {RIS} Optimization.
arXiv:2212.07333
(2022)
\end{botherref}
\endbibitem

\bibitem{tranter2017}
\begin{barticle}
\bauthor{\bsnm{Tranter}, \binits{J.}},
\bauthor{\bsnm{Sidiropoulos}, \binits{N.D.}},
\bauthor{\bsnm{Fu}, \binits{X.}},
\bauthor{\bsnm{Swami}, \binits{A.}}:
\batitle{Fast unit-modulus least squares with applications in beamforming}.
\bjtitle{IEEE Trans. Signal Process.}
\bvolume{65}(\bissue{11}),
\bfpage{2875}--\blpage{2887}
(\byear{2017}).
doi:\doiurl{10.1109/TSP.2017.2666774}
\end{barticle}
\endbibitem

\bibitem{Huang_GLOBECOM_2019}
\begin{bchapter}
\bauthor{\bsnm{{Huang}}, \binits{C.}},
\bauthor{\bsnm{{Alexandropoulos}}, \binits{G.C.}},
\bauthor{\bsnm{{Zappone}}, \binits{A.}},
\bauthor{\bsnm{{Debbah}}, \binits{M.}},
\bauthor{\bsnm{{Yuen}}, \binits{C.}}:
\bctitle{Energy efficient multi-user {MISO} communication using low resolution
  large intelligent surfaces}.
In: \bbtitle{Proc. IEEE Global Communications Conference (GLOBECOM)}
(\byear{2018})
\end{bchapter}
\endbibitem

\bibitem{basar2019wireless}
\begin{barticle}
\bauthor{\bsnm{Basar}, \binits{E.}},
\bauthor{\bsnm{Di~Renzo}, \binits{M.}},
\bauthor{\bsnm{De~Rosny}, \binits{J.}},
\bauthor{\bsnm{Debbah}, \binits{M.}},
\bauthor{\bsnm{Alouini}, \binits{M.-S.}},
\bauthor{\bsnm{Zhang}, \binits{R.}}:
\batitle{Wireless communications through reconfigurable intelligent surfaces}.
\bjtitle{IEEE access}
\bvolume{7},
\bfpage{116753}--\blpage{116773}
(\byear{2019})
\end{barticle}
\endbibitem

\bibitem{AbuShaban_2021}
\begin{bchapter}
\bauthor{\bsnm{Abu-Shaban}, \binits{Z.}},
\bauthor{\bsnm{Keykhosravi}, \binits{K.}},
\bauthor{\bsnm{Keskin}, \binits{M.F.}},
\bauthor{\bsnm{Alexandropoulos}, \binits{G.C.}},
\bauthor{\bsnm{Seco-Granados}, \binits{G.}},
\bauthor{\bsnm{Wymeersch}, \binits{H.}}:
\bctitle{Near-field localization with a reconfigurable intelligent surface
  acting as lens}.
In: \bbtitle{Proc. IEEE International Conference on Communications (ICC)}
(\byear{2021})
\end{bchapter}
\endbibitem

\bibitem{li2007range}
\begin{barticle}
\bauthor{\bsnm{Li}, \binits{J.}},
\bauthor{\bsnm{Xu}, \binits{L.}},
\bauthor{\bsnm{Stoica}, \binits{P.}},
\bauthor{\bsnm{Forsythe}, \binits{K.W.}},
\bauthor{\bsnm{Bliss}, \binits{D.W.}}:
\batitle{Range compression and waveform optimization for {MIMO} radar: A
  {Cram{\'e}r--Rao} bound based study}.
\bjtitle{IEEE Trans. Signal Process.}
\bvolume{56}(\bissue{1}),
\bfpage{218}--\blpage{232}
(\byear{2008}).
doi:\doiurl{10.1109/TSP.2007.901653}
\end{barticle}
\endbibitem

\bibitem{keskin_optimal_2021}
\begin{barticle}
\bauthor{\bsnm{Keskin}, \binits{F.}},
\bauthor{\bsnm{Jiang}, \binits{F.}},
\bauthor{\bsnm{Munier}, \binits{F.}},
\bauthor{\bsnm{Seco-Granados}, \binits{G.}},
\bauthor{\bsnm{Wymeersch}, \binits{H.}}:
\batitle{Optimal spatial signal design for mmwave positioning under imperfect
  synchronization}.
\bjtitle{IEEE Trans. Veh. Technol.}
(\byear{2022}).
doi:\doiurl{10.1109/TVT.2022.3149974}
\end{barticle}
\endbibitem

\bibitem{zhang2018multibeam}
\begin{barticle}
\bauthor{\bsnm{Zhang}, \binits{J.A.}},
\bauthor{\bsnm{Huang}, \binits{X.}},
\bauthor{\bsnm{Guo}, \binits{Y.J.}},
\bauthor{\bsnm{Yuan}, \binits{J.}},
\bauthor{\bsnm{Heath~Jr.}, \binits{R.W.}}:
\batitle{Multibeam for joint communication and radar sensing using steerable
  analog antenna arrays}.
\bjtitle{IEEE Trans. Veh. Technol.}
\bvolume{68}(\bissue{1}),
\bfpage{671}--\blpage{685}
(\byear{2019})
\end{barticle}
\endbibitem

\bibitem{kay_fundamentals}
\begin{botherref}
\oauthor{\bsnm{Kay}, \binits{S.M.}}:
Fundamentals of {Statistical} {Signal} {Processing}: {Practical} Algorithm
  Development.
Prentice-Hall PTR, 2013
\end{botherref}
\endbibitem

\bibitem{vanTrees}
\begin{botherref}
\oauthor{\bsnm{Van~Trees}, \binits{H.L.}}:
Detection, {Estimation}, and {Modulation} {Theory}, {Part} {I}: {Detection},
  {Estimation}, and {Linear} {Modulation} {Theory}
vol. pt. 1.
Wiley \& Sons, Ltd, 2004
\end{botherref}
\endbibitem

\bibitem{boyd2004convex}
\begin{botherref}
\oauthor{\bsnm{Boyd}, \binits{S.}},
\oauthor{\bsnm{Vandenberghe}, \binits{L.}}:
Convex Optimization.
Cambridge university press, 2004
\end{botherref}
\endbibitem

\bibitem{garcia_optimal_2018}
\begin{barticle}
\bauthor{\bsnm{Garcia}, \binits{N.}},
\bauthor{\bsnm{Wymeersch}, \binits{H.}},
\bauthor{\bsnm{Slock}, \binits{D.}}:
\batitle{Optimal {Precoders} for {Tracking} the {AoD} and {AoA} of a mm-{Wave}
  {Path}}.
\bjtitle{IEEE Trans. Signal Process.}
\bvolume{66}(\bissue{21}),
\bfpage{5718}--\blpage{5729}
(\byear{2018}).
doi:\doiurl{10.1109/TSP.2018.2870368}
\end{barticle}
\endbibitem

\bibitem{4359542}
\begin{barticle}
\bauthor{\bsnm{Li}, \binits{J.}},
\bauthor{\bsnm{Xu}, \binits{L.}},
\bauthor{\bsnm{Stoica}, \binits{P.}},
\bauthor{\bsnm{Forsythe}, \binits{K.W.}},
\bauthor{\bsnm{Bliss}, \binits{D.W.}}:
\batitle{Range compression and waveform optimization for {MIMO} radar: A
  {Cram{\'e}r--Rao} bound based study}.
\bjtitle{IEEE Transactions on Signal Processing}
\bvolume{56}(\bissue{1}),
\bfpage{218}--\blpage{232}
(\byear{2008}).
doi:\doiurl{10.1109/TSP.2007.901653}
\end{barticle}
\endbibitem

\bibitem{cvx}
\begin{botherref}
\oauthor{\bsnm{Grant}, \binits{M.}},
\oauthor{\bsnm{Boyd}, \binits{S.}}:
{CVX}: Matlab Software for Disciplined Convex Programming, version 2.1.
\url{http://cvxr.com/cvx}
(2014)
\end{botherref}
\endbibitem

\end{thebibliography}

\newcommand{\BMCxmlcomment}[1]{}

\BMCxmlcomment{

<refgrp>

<bibl id="B1">
  <title><p>Reconfigurable intelligent surfaces for energy efficiency in
  wireless communication</p></title>
  <aug>
    <au><snm>Huang</snm><fnm>C.</fnm></au>
    <au><snm>Zappone</snm><fnm>A.</fnm></au>
    <au><snm>Alexandropoulos</snm><fnm>G. C.</fnm></au>
    <au><snm>Debbah</snm><fnm>M.</fnm></au>
    <au><snm>Yuen</snm><fnm>C.</fnm></au>
  </aug>
  <source>IEEE Trans. Wireless Commun.</source>
  <publisher>IEEE</publisher>
  <pubdate>2019</pubdate>
  <volume>18</volume>
  <issue>8</issue>
  <fpage>4157</fpage>
  <lpage>-4170</lpage>
</bibl>

<bibl id="B2">
  <title><p>Intelligent Reflecting Surface Enhanced Wireless Network via Joint
  Active and Passive Beamforming</p></title>
  <aug>
    <au><snm>Wu</snm><fnm>Q.</fnm></au>
    <au><snm>Zhang</snm><fnm>R.</fnm></au>
  </aug>
  <source>IEEE Trans. Wireless Commun.</source>
  <pubdate>2019</pubdate>
  <volume>18</volume>
  <issue>11</issue>
  <fpage>5394</fpage>
  <lpage>5409</lpage>
</bibl>

<bibl id="B3">
  <title><p>Reconfigurable, intelligent, and sustainable wireless environments
  for {6G} smart connectivity</p></title>
  <aug>
    <au><snm>Calvanese Strinati</snm><fnm>E.</fnm></au>
    <au><snm>Alexandropoulos</snm><fnm>G. C.</fnm></au>
    <au><snm>Wymeersch</snm><fnm>H.</fnm></au>
    <au><snm>Denis</snm><fnm>B.</fnm></au>
    <au><snm>Sciancalepore</snm><fnm>V.</fnm></au>
    <au><snm>D'Errico</snm><fnm>R.</fnm></au>
    <au><snm>Clemente</snm><fnm>A.</fnm></au>
    <au><snm>Phan Huy</snm><fnm>D. T.</fnm></au>
    <au><snm>Carvalho</snm><fnm>ED</fnm></au>
    <au><snm>Popovski</snm><fnm>P.</fnm></au>
  </aug>
  <source>IEEE Commun. Mag.</source>
  <pubdate>2021</pubdate>
  <volume>59</volume>
  <issue>10</issue>
  <fpage>99</fpage>
  <lpage>-105</lpage>
</bibl>

<bibl id="B4">
  <title><p>Leveraging RIS-Enabled Smart Signal Propagation for Solving
  Infeasible Localization Problems</p></title>
  <aug>
    <au><snm>Keykhosravi</snm><fnm>K</fnm></au>
    <au><snm>Denis</snm><fnm>B</fnm></au>
    <au><snm>Alexandropoulos</snm><fnm>GC</fnm></au>
    <au><snm>He</snm><fnm>ZS</fnm></au>
    <au><snm>Albanese</snm><fnm>A</fnm></au>
    <au><snm>Sciancalepore</snm><fnm>V</fnm></au>
    <au><snm>Wymeersch</snm><fnm>H</fnm></au>
  </aug>
  <pubdate>2022</pubdate>
</bibl>

<bibl id="B5">
  <title><p>Hybrid Beamforming for Massive {MIMO}: {A} Survey</p></title>
  <aug>
    <au><snm>Molisch</snm><fnm>A. F.</fnm></au>
    <au><snm>Ratnam</snm><fnm>V. V.</fnm></au>
    <au><snm>Han</snm><fnm>S.</fnm></au>
    <au><snm>Li</snm><fnm>Z.</fnm></au>
    <au><snm>Nguyen</snm><fnm>S. L. H.</fnm></au>
    <au><snm>Li</snm><fnm>L.</fnm></au>
    <au><snm>Haneda</snm><fnm>K.</fnm></au>
  </aug>
  <source>{IEEE} {Commun.} {M}ag.</source>
  <pubdate>2017</pubdate>
  <volume>55</volume>
  <issue>9</issue>
  <fpage>134</fpage>
  <lpage>-141</lpage>
</bibl>

<bibl id="B6">
  <title><p>Reconfigurable intelligent surfaces for rich scattering wireless
  communications: {R}ecent experiments, challenges, and
  opportunities</p></title>
  <aug>
    <au><snm>Alexandropoulos</snm><fnm>G. C.</fnm></au>
    <au><snm>Shlezinger</snm><fnm>N.</fnm></au>
    <au><snm>Hougne</snm><fnm>P.</fnm></au>
  </aug>
  <source>IEEE Commun. Mag.</source>
  <pubdate>2021</pubdate>
  <volume>59</volume>
  <issue>6</issue>
  <fpage>28</fpage>
  <lpage>-34</lpage>
</bibl>

<bibl id="B7">
  <title><p>Intelligent Reflecting Surface: {P}ractical Phase Shift Model and
  Beamforming Optimization</p></title>
  <aug>
    <au><snm>Abeywickrama</snm><fnm>S.</fnm></au>
    <au><snm>Zhang</snm><fnm>R.</fnm></au>
    <au><snm>Wu</snm><fnm>Q.</fnm></au>
    <au><snm>Yue</snm><fnm>C.</fnm></au>
  </aug>
  <source>IEEE Trans. Commun.</source>
  <pubdate>2020</pubdate>
  <volume>68</volume>
  <issue>9</issue>
  <fpage>5849</fpage>
  <lpage>-5863</lpage>
</bibl>

<bibl id="B8">
  <title><p>Active Reconfigurable Intelligent Surface-Aided Wireless
  Communications</p></title>
  <aug>
    <au><snm>Long</snm><fnm>R.</fnm></au>
    <au><snm>Liang</snm><fnm>Y. C.</fnm></au>
    <au><snm>Pei</snm><fnm>Y.</fnm></au>
    <au><snm>Larsson</snm><fnm>E. G.</fnm></au>
  </aug>
  <source>IEEE Trans. Wireless Commun.</source>
  <pubdate>2021</pubdate>
  <volume>20</volume>
  <issue>8</issue>
  <fpage>4962</fpage>
  <lpage>-4975</lpage>
</bibl>

<bibl id="B9">
  <title><p>Capacity Degradation with Modeling Hardware Impairment in Large
  Intelligent Surface</p></title>
  <aug>
    <au><snm>Hu</snm><fnm>S.</fnm></au>
    <au><snm>Rusek</snm><fnm>F.</fnm></au>
    <au><snm>Edfors</snm><fnm>O.</fnm></au>
  </aug>
  <source>Proc. IEEE Global Communications Conference (GLOBECOM)</source>
  <pubdate>2018</pubdate>
</bibl>

<bibl id="B10">
  <title><p>Beamforming optimization for {IRS}-aided communications with
  transceiver hardware impairments</p></title>
  <aug>
    <au><snm>Shen</snm><fnm>H</fnm></au>
    <au><snm>Xu</snm><fnm>W</fnm></au>
    <au><snm>Gong</snm><fnm>S</fnm></au>
    <au><snm>Zhao</snm><fnm>C</fnm></au>
    <au><snm>Ng</snm><fnm>DWK</fnm></au>
  </aug>
  <source>IEEE Trans. Commun.</source>
  <publisher>IEEE</publisher>
  <pubdate>2020</pubdate>
  <volume>69</volume>
  <issue>2</issue>
  <fpage>1214</fpage>
  <lpage>-1227</lpage>
</bibl>

<bibl id="B11">
  <title><p>Arbitrary Beam Pattern Approximation via RISs with Measured Element
  Responses</p></title>
  <aug>
    <au><snm>Rahal</snm><fnm>M</fnm></au>
    <au><snm>Denis</snm><fnm>B</fnm></au>
    <au><snm>Keykhosravi</snm><fnm>K</fnm></au>
    <au><snm>Keskin</snm><fnm>MF</fnm></au>
    <au><snm>Uguen</snm><fnm>B</fnm></au>
    <au><snm>Alexandropoulos</snm><fnm>GC</fnm></au>
    <au><snm>Wymeersch</snm><fnm>H</fnm></au>
  </aug>
  <source>Joint European Conference on Networks and Communications \& 6G Summit
  (EuCNC/6G Summit)</source>
  <pubdate>2022</pubdate>
  <fpage>506</fpage>
  <lpage>511</lpage>
</bibl>

<bibl id="B12">
  <title><p>Reconfigurable Intelligent Surfaces for Localization: Position and
  Orientation Error Bounds</p></title>
  <aug>
    <au><snm>Elzanaty</snm><fnm>A</fnm></au>
    <au><snm>Guerra</snm><fnm>A</fnm></au>
    <au><snm>Guidi</snm><fnm>F</fnm></au>
    <au><snm>Alouini</snm><fnm>MS</fnm></au>
  </aug>
  <source>IEEE Transactions on Signal Processing</source>
  <pubdate>2021</pubdate>
  <volume>69</volume>
  <fpage>5386</fpage>
  <lpage>5402</lpage>
</bibl>

<bibl id="B13">
  <title><p>Constrained RIS Phase Profile Optimization and Time Sharing for
  Near-field Localization</p></title>
  <aug>
    <au><snm>Rahal</snm><fnm>M</fnm></au>
    <au><snm>Denis</snm><fnm>B</fnm></au>
    <au><snm>Keykhosravi</snm><fnm>K</fnm></au>
    <au><snm>Keskin</snm><fnm>MF</fnm></au>
    <au><snm>Uguen</snm><fnm>B</fnm></au>
    <au><snm>Wymeersch</snm><fnm>H</fnm></au>
  </aug>
  <source>IEEE 95th Vehicular Technology Conference (VTC-Spring)</source>
  <pubdate>2022</pubdate>
  <fpage>1</fpage>
  <lpage>6</lpage>
</bibl>

<bibl id="B14">
  <title><p>{SISO} {RIS}-enabled joint {3D} downlink localization and
  synchronization</p></title>
  <aug>
    <au><snm>Keykhosravi</snm><fnm>K</fnm></au>
    <au><snm>Keskin</snm><fnm>MF</fnm></au>
    <au><snm>Seco Granados</snm><fnm>G</fnm></au>
    <au><snm>Wymeersch</snm><fnm>H</fnm></au>
  </aug>
  <source>Proc. IEEE International Conference on Communications (ICC)</source>
  <pubdate>2021</pubdate>
</bibl>

<bibl id="B15">
  <title><p>{RIS}-Enabled Localization Continuity Under Near-Field
  Conditions</p></title>
  <aug>
    <au><snm>Rahal</snm><fnm>M</fnm></au>
    <au><snm>Denis</snm><fnm>B</fnm></au>
    <au><snm>Keykhosravi</snm><fnm>K</fnm></au>
    <au><snm>Uguen</snm><fnm>B</fnm></au>
    <au><snm>Wymeersch</snm><fnm>H</fnm></au>
  </aug>
  <source>Proc. IEEE 22nd International Workshop on Signal Processing Advances
  in Wireless Communications (SPAWC)</source>
  <pubdate>2021</pubdate>
</bibl>

<bibl id="B16">
  <title><p>A Prototype of Reconfigurable Intelligent Surface with Continuous
  Control of the Reflection Phase</p></title>
  <aug>
    <au><snm>Fara</snm><fnm>R</fnm></au>
    <au><snm>Ratajczak</snm><fnm>P</fnm></au>
    <au><snm>Huy</snm><fnm>DTP</fnm></au>
    <au><snm>Ourir</snm><fnm>A</fnm></au>
    <au><snm>Di Renzo</snm><fnm>M</fnm></au>
    <au><snm>De Rosny</snm><fnm>J</fnm></au>
  </aug>
  <pubdate>2021</pubdate>
</bibl>

<bibl id="B17">
  <title><p>Circularly-Polarized Reconfigurable Transmitarray in {K}a-Band With
  Beam Scanning and Polarization Switching Capabilities</p></title>
  <aug>
    <au><snm>Di Palma</snm><fnm>L</fnm></au>
    <au><snm>Clemente</snm><fnm>A</fnm></au>
    <au><snm>Dussopt</snm><fnm>L</fnm></au>
    <au><snm>Sauleau</snm><fnm>R</fnm></au>
    <au><snm>Potier</snm><fnm>P</fnm></au>
    <au><snm>Pouliguen</snm><fnm>P</fnm></au>
  </aug>
  <source>IEEE Trans. Antennas Prop.</source>
  <pubdate>2017</pubdate>
  <volume>65</volume>
  <issue>2</issue>
  <fpage>529</fpage>
  <lpage>540</lpage>
</bibl>

<bibl id="B18">
  <title><p>A Reconfigurable Intelligent Surface at mmWave Based on a Binary
  Phase Tunable Metasurface</p></title>
  <aug>
    <au><snm>Gros</snm><fnm>JB</fnm></au>
    <au><snm>Popov</snm><fnm>V</fnm></au>
    <au><snm>Odit</snm><fnm>MA</fnm></au>
    <au><snm>Lenets</snm><fnm>V</fnm></au>
    <au><snm>Lerosey</snm><fnm>G</fnm></au>
  </aug>
  <source>IEEE Open Journal of the Communications Society</source>
  <pubdate>2021</pubdate>
  <volume>2</volume>
  <fpage>1055</fpage>
  <lpage>1064</lpage>
</bibl>

<bibl id="B19">
  <title><p>Overview of generalized monopulse estimation</p></title>
  <aug>
    <au><snm>Nickel</snm><fnm>U</fnm></au>
  </aug>
  <source>IEEE Trans. Aerosp. Electron. Syst.</source>
  <publisher>IEEE</publisher>
  <pubdate>2006</pubdate>
  <volume>21</volume>
  <issue>6</issue>
  <fpage>27</fpage>
  <lpage>-56</lpage>
</bibl>

<bibl id="B20">
  <title><p>Beamformer architectures for active phased-array radar
  antennas</p></title>
  <aug>
    <au><snm>Agrawal</snm><fnm>A.K.</fnm></au>
    <au><snm>Holzman</snm><fnm>E.L.</fnm></au>
  </aug>
  <source>IEEE Trans. Antennas Prop.</source>
  <pubdate>1999</pubdate>
  <volume>47</volume>
  <issue>3</issue>
  <fpage>432</fpage>
  <lpage>442</lpage>
</bibl>

<bibl id="B21">
  <title><p>Benefits of Digital Phased Array Radars</p></title>
  <aug>
    <au><snm>Talisa</snm><fnm>SH</fnm></au>
    <au><snm>O'Haver</snm><fnm>KW</fnm></au>
    <au><snm>Comberiate</snm><fnm>TM</fnm></au>
    <au><snm>Sharp</snm><fnm>MD</fnm></au>
    <au><snm>Somerlock</snm><fnm>OF</fnm></au>
  </aug>
  <source>Proc. IEEE</source>
  <pubdate>2016</pubdate>
  <volume>104</volume>
  <issue>3</issue>
  <fpage>530</fpage>
  <lpage>543</lpage>
</bibl>

<bibl id="B22">
  <title><p>Intelligent Reflecting Surfaces: Sum-Rate Optimization Based on
  Statistical Position Information</p></title>
  <aug>
    <au><snm>Abrardo</snm><fnm>A</fnm></au>
    <au><snm>Dardari</snm><fnm>D</fnm></au>
    <au><snm>Di Renzo</snm><fnm>M</fnm></au>
  </aug>
  <source>IEEE Transactions on Communications, July</source>
  <pubdate>2021</pubdate>
</bibl>

<bibl id="B23">
  <title><p>Location information aided multiple intelligent reflecting surface
  systems</p></title>
  <aug>
    <au><snm>Hu</snm><fnm>X</fnm></au>
    <au><snm>Zhong</snm><fnm>C</fnm></au>
    <au><snm>Zhang</snm><fnm>Y</fnm></au>
    <au><snm>Chen</snm><fnm>X</fnm></au>
    <au><snm>Zhang</snm><fnm>Z</fnm></au>
  </aug>
  <source>IEEE Transactions on Communications</source>
  <publisher>IEEE</publisher>
  <pubdate>2020</pubdate>
  <volume>68</volume>
  <issue>12</issue>
  <fpage>7948</fpage>
  <lpage>-7962</lpage>
</bibl>

<bibl id="B24">
  <title><p>Beyond {5G} {Wireless} {Localization} with {Reconfigurable}
  {Intelligent} {Surfaces}</p></title>
  <aug>
    <au><snm>Wymeersch</snm><fnm>H</fnm></au>
    <au><snm>Denis</snm><fnm>B</fnm></au>
  </aug>
  <source>{IEEE} {International} {Conference} on {Communications}
  ({ICC})</source>
  <pubdate>2020</pubdate>
  <fpage>1</fpage>
  <lpage>-6</lpage>
</bibl>

<bibl id="B25">
  <title><p>Compressive Near-Field Localization for Multipath RIS-Aided
  Environments</p></title>
  <aug>
    <au><snm>Rinchi</snm><fnm>O</fnm></au>
    <au><snm>Elzanaty</snm><fnm>A</fnm></au>
    <au><snm>Alouini</snm><fnm>MS</fnm></au>
  </aug>
  <source>IEEE Communications Letters</source>
  <pubdate>2022</pubdate>
  <volume>26</volume>
  <issue>6</issue>
  <fpage>1268</fpage>
  <lpage>1272</lpage>
</bibl>

<bibl id="B26">
  <title><p>RIS-aided User Tracking in Near-Field MIMO Systems: Joint Precoding
  Design and RIS Optimization</p></title>
  <aug>
    <au><snm>Palmucci</snm><fnm>S</fnm></au>
    <au><snm>Guerra</snm><fnm>A</fnm></au>
    <au><snm>Abrardo</snm><fnm>A</fnm></au>
    <au><snm>Dardari</snm><fnm>D</fnm></au>
  </aug>
  <pubdate>2022</pubdate>
</bibl>

<bibl id="B27">
  <title><p>RIS-Aided Near-Field Localization and Channel Estimation for the
  Sub-Terahertz System</p></title>
  <aug>
    <au><snm>Pan</snm><fnm>Y</fnm></au>
    <au><snm>Pan</snm><fnm>C</fnm></au>
    <au><snm>Jin</snm><fnm>S</fnm></au>
    <au><snm>Wang</snm><fnm>J</fnm></au>
  </aug>
  <publisher>arXiv</publisher>
  <pubdate>2022</pubdate>
</bibl>

<bibl id="B28">
  <title><p>Bi-Static Sensing for Near-Field RIS Localization</p></title>
  <aug>
    <au><snm>Ghazalian</snm><fnm>R</fnm></au>
    <au><snm>Keykhosravi</snm><fnm>K</fnm></au>
    <au><snm>Chen</snm><fnm>H</fnm></au>
    <au><snm>Wymeersch</snm><fnm>H</fnm></au>
    <au><snm>Jäntti</snm><fnm>R</fnm></au>
  </aug>
  <source>IEEE Global Communications Conference (GLOBECOM)</source>
  <pubdate>2022</pubdate>
  <fpage>6457</fpage>
  <lpage>6462</lpage>
</bibl>

<bibl id="B29">
  <title><p>On the Study of Reconfigurable Intelligent Surfaces in the
  Near-Field Region</p></title>
  <aug>
    <au><snm>Mei</snm><fnm>P</fnm></au>
    <au><snm>Cai</snm><fnm>Y</fnm></au>
    <au><snm>Zhao</snm><fnm>K</fnm></au>
    <au><snm>Ying</snm><fnm>Z</fnm></au>
    <au><snm>Pedersen</snm><fnm>GF</fnm></au>
    <au><snm>Lin</snm><fnm>XQ</fnm></au>
    <au><snm>Zhang</snm><fnm>S</fnm></au>
  </aug>
  <source>IEEE Transactions on Antennas and Propagation</source>
  <pubdate>2022</pubdate>
  <volume>70</volume>
  <issue>10</issue>
  <fpage>8718</fpage>
  <lpage>8728</lpage>
</bibl>

<bibl id="B30">
  <title><p>Near-Field Hierarchical Beam Management for {RIS}-Enabled
  Millimeter Wave Multi-Antenna Systems</p></title>
  <aug>
    <au><snm>Alexandropoulos</snm><fnm>GC</fnm></au>
    <au><snm>Jamali</snm><fnm>V</fnm></au>
    <au><snm>Schober</snm><fnm>R</fnm></au>
    <au><snm>Poor</snm><fnm>HV</fnm></au>
  </aug>
  <source>Proc. IEEE SAM</source>
  <pubdate>2022</pubdate>
  <fpage>460</fpage>
  <lpage>464</lpage>
</bibl>

<bibl id="B31">
  <title><p>Low-overhead Beam Training Scheme for Extremely Large-Scale {RIS}
  in Near-field</p></title>
  <aug>
    <au><snm>Liu</snm><fnm>W</fnm></au>
    <au><snm>Pan</snm><fnm>C</fnm></au>
    <au><snm>Ren</snm><fnm>H</fnm></au>
    <au><snm>Shu</snm><fnm>F</fnm></au>
    <au><snm>Jin</snm><fnm>S</fnm></au>
    <au><snm>Wang</snm><fnm>J</fnm></au>
  </aug>
  <pubdate>2022</pubdate>
  <note>arXiv:2211.15910</note>
</bibl>

<bibl id="B32">
  <title><p>Location-aware Channel Estimation for {RIS}-aided {mmWave} {MIMO}
  Systems via Atomic Norm Minimization</p></title>
  <aug>
    <au><snm>Chung</snm><fnm>H</fnm></au>
    <au><snm>Kim</snm><fnm>S</fnm></au>
  </aug>
  <pubdate>2021</pubdate>
  <note>arXiv:2107.09222</note>
</bibl>

<bibl id="B33">
  <title><p>Differential Data-Aided Beam Training for {RIS}-Empowered
  Multi-Antenna Communications</p></title>
  <aug>
    <au><snm>Chen Hu</snm><fnm>K</fnm></au>
    <au><snm>Alexandropoulos</snm><fnm>GC</fnm></au>
    <au><snm>Armada</snm><fnm>Ag</fnm></au>
  </aug>
  <source>IEEE Access</source>
  <pubdate>2022</pubdate>
  <volume>10</volume>
  <fpage>113200</fpage>
  <lpage>113213</lpage>
</bibl>

<bibl id="B34">
  <title><p>{RIS}-aided User Tracking in Near-Field {MIMO} Systems: {Joint}
  Precoding Design and {RIS} Optimization</p></title>
  <aug>
    <au><snm>Palmucci</snm><fnm>S</fnm></au>
    <au><snm>Guerra</snm><fnm>A</fnm></au>
    <au><snm>Abrardo</snm><fnm>A</fnm></au>
    <au><snm>Dardari</snm><fnm>D</fnm></au>
  </aug>
  <pubdate>2022</pubdate>
  <note>arXiv:2212.07333</note>
</bibl>

<bibl id="B35">
  <title><p>Fast Unit-Modulus Least Squares With Applications in
  Beamforming</p></title>
  <aug>
    <au><snm>Tranter</snm><fnm>J</fnm></au>
    <au><snm>Sidiropoulos</snm><fnm>ND</fnm></au>
    <au><snm>Fu</snm><fnm>X</fnm></au>
    <au><snm>Swami</snm><fnm>A</fnm></au>
  </aug>
  <source>IEEE Trans. Signal Process.</source>
  <pubdate>2017</pubdate>
  <volume>65</volume>
  <issue>11</issue>
  <fpage>2875</fpage>
  <lpage>2887</lpage>
</bibl>

<bibl id="B36">
  <title><p>Energy Efficient Multi-User {MISO} Communication Using Low
  Resolution Large Intelligent Surfaces</p></title>
  <aug>
    <au><snm>{Huang}</snm><fnm>C.</fnm></au>
    <au><snm>{Alexandropoulos}</snm><fnm>G. C.</fnm></au>
    <au><snm>{Zappone}</snm><fnm>A.</fnm></au>
    <au><snm>{Debbah}</snm><fnm>M.</fnm></au>
    <au><snm>{Yuen}</snm><fnm>C.</fnm></au>
  </aug>
  <source>Proc. IEEE Global Communications Conference (GLOBECOM)</source>
  <pubdate>2018</pubdate>
</bibl>

<bibl id="B37">
  <title><p>Wireless communications through reconfigurable intelligent
  surfaces</p></title>
  <aug>
    <au><snm>Basar</snm><fnm>E</fnm></au>
    <au><snm>Di Renzo</snm><fnm>M</fnm></au>
    <au><snm>De Rosny</snm><fnm>J</fnm></au>
    <au><snm>Debbah</snm><fnm>M</fnm></au>
    <au><snm>Alouini</snm><fnm>MS</fnm></au>
    <au><snm>Zhang</snm><fnm>R</fnm></au>
  </aug>
  <source>IEEE access</source>
  <publisher>IEEE</publisher>
  <pubdate>2019</pubdate>
  <volume>7</volume>
  <fpage>116753</fpage>
  <lpage>-116773</lpage>
</bibl>

<bibl id="B38">
  <title><p>Near-field Localization with a Reconfigurable Intelligent Surface
  Acting as Lens</p></title>
  <aug>
    <au><snm>Abu Shaban</snm><fnm>Z</fnm></au>
    <au><snm>Keykhosravi</snm><fnm>K</fnm></au>
    <au><snm>Keskin</snm><fnm>MF</fnm></au>
    <au><snm>Alexandropoulos</snm><fnm>GC</fnm></au>
    <au><snm>Seco Granados</snm><fnm>G</fnm></au>
    <au><snm>Wymeersch</snm><fnm>H</fnm></au>
  </aug>
  <source>Proc. IEEE International Conference on Communications (ICC)</source>
  <pubdate>2021</pubdate>
</bibl>

<bibl id="B39">
  <title><p>Range Compression and Waveform Optimization for {MIMO} Radar: A
  {Cram{\'e}r--Rao} Bound Based Study</p></title>
  <aug>
    <au><snm>Li</snm><fnm>J</fnm></au>
    <au><snm>Xu</snm><fnm>L</fnm></au>
    <au><snm>Stoica</snm><fnm>P</fnm></au>
    <au><snm>Forsythe</snm><fnm>KW</fnm></au>
    <au><snm>Bliss</snm><fnm>DW</fnm></au>
  </aug>
  <source>IEEE Trans. Signal Process.</source>
  <pubdate>2008</pubdate>
  <volume>56</volume>
  <issue>1</issue>
  <fpage>218</fpage>
  <lpage>232</lpage>
</bibl>

<bibl id="B40">
  <title><p>Optimal Spatial Signal Design for mmWave Positioning under
  Imperfect Synchronization</p></title>
  <aug>
    <au><snm>Keskin</snm><fnm>F</fnm></au>
    <au><snm>Jiang</snm><fnm>F</fnm></au>
    <au><snm>Munier</snm><fnm>F</fnm></au>
    <au><snm>Seco Granados</snm><fnm>G</fnm></au>
    <au><snm>Wymeersch</snm><fnm>H</fnm></au>
  </aug>
  <source>IEEE Trans. Veh. Technol.</source>
  <pubdate>2022</pubdate>
</bibl>

<bibl id="B41">
  <title><p>Multibeam for joint communication and radar sensing using steerable
  analog antenna arrays</p></title>
  <aug>
    <au><snm>Zhang</snm><fnm>JA</fnm></au>
    <au><snm>Huang</snm><fnm>X</fnm></au>
    <au><snm>Guo</snm><fnm>YJ</fnm></au>
    <au><snm>Yuan</snm><fnm>J</fnm></au>
    <au><snm>Heath Jr.</snm><fnm>R. W</fnm></au>
  </aug>
  <source>IEEE Trans. Veh. Technol.</source>
  <publisher>IEEE</publisher>
  <pubdate>2019</pubdate>
  <volume>68</volume>
  <issue>1</issue>
  <fpage>671</fpage>
  <lpage>-685</lpage>
</bibl>

<bibl id="B42">
  <title><p>Fundamentals of {Statistical} {Signal} {Processing}: {Practical}
  algorithm development</p></title>
  <aug>
    <au><snm>Kay</snm><fnm>S.M.</fnm></au>
  </aug>
  <publisher>Prentice-Hall PTR, 2013</publisher>
</bibl>

<bibl id="B43">
  <title><p>Detection, {Estimation}, and {Modulation} {Theory}, {Part} {I}:
  {Detection}, {Estimation}, and {Linear} {Modulation} {Theory}</p></title>
  <aug>
    <au><snm>Van Trees</snm><fnm>H.L.</fnm></au>
  </aug>
  <publisher>Wiley \& Sons, Ltd, 2004</publisher>
  <issue>pt. 1</issue>
</bibl>

<bibl id="B44">
  <title><p>Convex optimization</p></title>
  <aug>
    <au><snm>Boyd</snm><fnm>S</fnm></au>
    <au><snm>Vandenberghe</snm><fnm>L</fnm></au>
  </aug>
  <publisher>Cambridge university press, 2004</publisher>
</bibl>

<bibl id="B45">
  <title><p>Optimal {Precoders} for {Tracking} the {AoD} and {AoA} of a
  mm-{Wave} {Path}</p></title>
  <aug>
    <au><snm>Garcia</snm><fnm>N</fnm></au>
    <au><snm>Wymeersch</snm><fnm>H</fnm></au>
    <au><snm>Slock</snm><fnm>D</fnm></au>
  </aug>
  <source>IEEE Trans. Signal Process.</source>
  <pubdate>2018</pubdate>
  <volume>66</volume>
  <issue>21</issue>
  <fpage>5718</fpage>
  <lpage>-5729</lpage>
</bibl>

<bibl id="B46">
  <title><p>Range Compression and Waveform Optimization for {MIMO} Radar: A
  {Cram{\'e}r--Rao} Bound Based Study</p></title>
  <aug>
    <au><snm>Li</snm><fnm>J</fnm></au>
    <au><snm>Xu</snm><fnm>L</fnm></au>
    <au><snm>Stoica</snm><fnm>P</fnm></au>
    <au><snm>Forsythe</snm><fnm>KW</fnm></au>
    <au><snm>Bliss</snm><fnm>DW</fnm></au>
  </aug>
  <source>IEEE Transactions on Signal Processing</source>
  <pubdate>2008</pubdate>
  <volume>56</volume>
  <issue>1</issue>
  <fpage>218</fpage>
  <lpage>232</lpage>
</bibl>

<bibl id="B47">
  <title><p>{CVX}: Matlab Software for Disciplined Convex Programming, version
  2.1</p></title>
  <aug>
    <au><snm>Grant</snm><fnm>M</fnm></au>
    <au><snm>Boyd</snm><fnm>S</fnm></au>
  </aug>
  <source>\url{http://cvxr.com/cvx}</source>
  <pubdate>2014</pubdate>
</bibl>

</refgrp>
} 



\begin{thebibliography}{6}
\ifx \bisbn   \undefined \def \bisbn  #1{ISBN #1}\fi
\ifx \binits  \undefined \def \binits#1{#1}\fi
\ifx \bauthor  \undefined \def \bauthor#1{#1}\fi
\ifx \batitle  \undefined \def \batitle#1{#1}\fi
\ifx \bjtitle  \undefined \def \bjtitle#1{#1}\fi
\ifx \bvolume  \undefined \def \bvolume#1{\textbf{#1}}\fi
\ifx \byear  \undefined \def \byear#1{#1}\fi
\ifx \bissue  \undefined \def \bissue#1{#1}\fi
\ifx \bfpage  \undefined \def \bfpage#1{#1}\fi
\ifx \blpage  \undefined \def \blpage #1{#1}\fi
\ifx \burl  \undefined \def \burl#1{\textsf{#1}}\fi
\ifx \doiurl  \undefined \def \doiurl#1{\textsf{#1}}\fi
\ifx \betal  \undefined \def \betal{\textit{et al.}}\fi
\ifx \binstitute  \undefined \def \binstitute#1{#1}\fi
\ifx \binstitutionaled  \undefined \def \binstitutionaled#1{#1}\fi
\ifx \bctitle  \undefined \def \bctitle#1{#1}\fi
\ifx \beditor  \undefined \def \beditor#1{#1}\fi
\ifx \bpublisher  \undefined \def \bpublisher#1{#1}\fi
\ifx \bbtitle  \undefined \def \bbtitle#1{#1}\fi
\ifx \bedition  \undefined \def \bedition#1{#1}\fi
\ifx \bseriesno  \undefined \def \bseriesno#1{#1}\fi
\ifx \blocation  \undefined \def \blocation#1{#1}\fi
\ifx \bsertitle  \undefined \def \bsertitle#1{#1}\fi
\ifx \bsnm \undefined \def \bsnm#1{#1}\fi
\ifx \bsuffix \undefined \def \bsuffix#1{#1}\fi
\ifx \bparticle \undefined \def \bparticle#1{#1}\fi
\ifx \barticle \undefined \def \barticle#1{#1}\fi
\ifx \bconfdate \undefined \def \bconfdate #1{#1}\fi
\ifx \botherref \undefined \def \botherref #1{#1}\fi
\ifx \url \undefined \def \url#1{\textsf{#1}}\fi
\ifx \bchapter \undefined \def \bchapter#1{#1}\fi
\ifx \bbook \undefined \def \bbook#1{#1}\fi
\ifx \bcomment \undefined \def \bcomment#1{#1}\fi
\ifx \oauthor \undefined \def \oauthor#1{#1}\fi
\ifx \citeauthoryear \undefined \def \citeauthoryear#1{#1}\fi
\ifx \endbibitem  \undefined \def \endbibitem {}\fi
\ifx \bconflocation  \undefined \def \bconflocation#1{#1}\fi
\ifx \arxivurl  \undefined \def \arxivurl#1{\textsf{#1}}\fi
\csname PreBibitemsHook\endcsname

\bibitem{koon}
\begin{barticle}
\bauthor{\bsnm{Koonin}, \binits{E.V.}},
\bauthor{\bsnm{Altschul}, \binits{S.F.}},
\bauthor{\bsnm{Bork}, \binits{P.}}:
\batitle{Brca1 protein products: functional motifs}.
\bjtitle{Nat. Genet.}
\bvolume{13},
\bfpage{266}--\blpage{267}
(\byear{1996})
\end{barticle}
\endbibitem

\bibitem{xjon}
\begin{bchapter}
\bauthor{\bsnm{Jones}, \binits{X.}}:
\bctitle{Zeolites and synthetic mechanisms}.
In: \beditor{\bsnm{Smith}, \binits{Y.}} (ed.)
\bbtitle{Proceedings of the First National Conference on Porous Sieves: 27-30
  June 1996; Baltimore},
pp. \bfpage{16}--\blpage{27}
(\byear{1996})
\end{bchapter}
\endbibitem

\bibitem{marg}
\begin{bbook}
\bauthor{\bsnm{Margulis}, \binits{L.}}:
\bbtitle{Origin of Eukaryotic Cells}.
\bpublisher{Yale University Press},
\blocation{New Haven}
(\byear{1970})
\end{bbook}
\endbibitem

\bibitem{schn}
\begin{bchapter}
\bauthor{\bsnm{Schnepf}, \binits{E.}}:
\bctitle{From prey via endosymbiont to plastids: comparative studies in
  dinoflagellates}.
In: \beditor{\bsnm{Lewin}, \binits{R.A.}} (ed.)
\bbtitle{Origins of Plastids},
\bedition{2nd} edn.,
pp. \bfpage{53}--\blpage{76}.
\bpublisher{Chapman and Hall},
\blocation{New York}
(\byear{1993})
\end{bchapter}
\endbibitem

\bibitem{koha}
\begin{botherref}
\oauthor{\bsnm{Kohavi}, \binits{R.}}:
Wrappers for performance enhancement and obvious decision graphs.
PhD thesis,
Stanford University, Computer Science Department
(1995)
\end{botherref}
\endbibitem

\bibitem{issnic}
\begin{botherref}
\oauthor{\bsnm{{ISSN International Centre}}}:
The ISSN register
(2006).
\url{http://www.issn.org}
Accessed Accessed 20 Feb 2007
\end{botherref}
\endbibitem

\end{thebibliography}

\newcommand{\BMCxmlcomment}[1]{}

\BMCxmlcomment{

<refgrp>

<bibl id="B1">
  <title><p>BRCA1 protein products: functional motifs</p></title>
  <aug>
    <au><snm>Koonin</snm><fnm>E V</fnm></au>
    <au><snm>Altschul</snm><fnm>S F</fnm></au>
    <au><snm>Bork</snm><fnm>P</fnm></au>
  </aug>
  <source>Nat. Genet.</source>
  <pubdate>1996</pubdate>
  <volume>13</volume>
  <fpage>266</fpage>
  <lpage>267</lpage>
</bibl>

<bibl id="B2">
  <title><p>Zeolites and synthetic mechanisms</p></title>
  <aug>
    <au><snm>Jones</snm><fnm>X</fnm></au>
  </aug>
  <source>Proceedings of the First National Conference on Porous Sieves: 27-30
  June 1996; Baltimore</source>
  <editor>Y Smith</editor>
  <pubdate>1996</pubdate>
  <fpage>16</fpage>
  <lpage>27</lpage>
</bibl>

<bibl id="B3">
  <title><p>Origin of Eukaryotic Cells</p></title>
  <aug>
    <au><snm>Margulis</snm><fnm>L</fnm></au>
  </aug>
  <publisher>New Haven: Yale University Press</publisher>
  <pubdate>1970</pubdate>
</bibl>

<bibl id="B4">
  <title><p>From prey via endosymbiont to plastids: comparative studies in
  dinoflagellates</p></title>
  <aug>
    <au><snm>Schnepf</snm><fnm>E</fnm></au>
  </aug>
  <source>Origins of Plastids</source>
  <publisher>New York: Chapman and Hall</publisher>
  <editor>R A Lewin</editor>
  <edition>2</edition>
  <pubdate>1993</pubdate>
  <fpage>53</fpage>
  <lpage>76</lpage>
</bibl>

<bibl id="B5">
  <title><p>Wrappers for performance enhancement and obvious decision
  graphs</p></title>
  <aug>
    <au><snm>Kohavi</snm><fnm>R</fnm></au>
  </aug>
  <source>PhD thesis</source>
  <publisher>Stanford University, Computer Science Department</publisher>
  <pubdate>1995</pubdate>
</bibl>

<bibl id="B6">
  <title><p>The ISSN register</p></title>
  <aug>
    <au><cnm>{ISSN International Centre}</cnm></au>
  </aug>
  <pubdate>2006</pubdate>
  <url>http://www.issn.org</url>
</bibl>

</refgrp>
} 
\end{document}